%% file: main.tex
\documentclass[sigconf]{acmart}

\copyrightyear{2023}
\acmYear{2023}
\setcopyright{acmlicensed}\acmConference[WWW '23]{Proceedings of the ACM Web Conference 2023}{May 1--5, 2023}{Austin, TX, USA}
\acmBooktitle{Proceedings of the ACM Web Conference 2023 (WWW '23), May 1--5, 2023, Austin, TX, USA}
\acmPrice{15.00}
\acmDOI{10.1145/3543507.3583267}
\acmISBN{978-1-4503-9416-1/23/04}



\input{macros.tex}

\begin{document}

\title[The Thin Ideology of Populist Advertising on Facebook]{The Thin Ideology of Populist Advertising on Facebook\\during the 2019 EU Elections}

\author{Arthur Capozzi}
\email{arthurtomasedward.capozzi@unito.it}
\authornote{After the first author, the author names are in alphabetical order.}
\orcid{0000-0002-1996-1800}
\affiliation{%
  \institution{Università degli Studi di Torino}
  \streetaddress{Via Pessinetto 12}
  \city{Turin}
  \country{Italy}
  \postcode{10149}
}

\author{\texorpdfstring{\nohyphens{Gianmarco~De~Francisci~Morales}}{Gianmarco De Francisci Morales}}
\email{gdfm@acm.org}
\orcid{0000-0002-2415-494X}
\affiliation{%
  \institution{CENTAI}
  \streetaddress{Corso Inghilterra 3}
  \city{Turin}
  \country{Italy}
  \postcode{10138}
}

\author{Yelena Mejova}
\email{yelenamejova@acm.org}
\orcid{0000-0001-5560-4109}
\affiliation{%
  \institution{ISI Foundation}
  \streetaddress{Via Chisola 5}
  \city{Turin}
  \country{Italy}
  \postcode{10126}
}

\author{Corrado Monti}
\email{corrado.monti@centai.eu}
\orcid{0000-0001-6846-5718}
\affiliation{%
  \institution{CENTAI}
  \streetaddress{Corso Inghilterra 3}
  \city{Turin}
  \country{Italy}
  \postcode{10138}
}

\author{Andr\'{e} Panisson}
\email{andre.panisson@centai.eu}
\orcid{0000-0002-3336-0374}
\affiliation{%
  \institution{CENTAI}
  \streetaddress{Corso Inghilterra 3}
  \city{Turin}
  \country{Italy}
  \postcode{10138}
}
\renewcommand{\shortauthors}{Capozzi, et al.}

\begin{abstract}
Social media has been an important tool in the expansion of the populist message, and it is thought to have contributed to the electoral success of populist parties in the past decade.
This study compares how populist parties advertised on Facebook during the 2019 European Parliamentary election.
In particular, we examine commonalities and differences in which audiences they reach and on which issues they focus.
By using data from Meta (previously Facebook) Ad Library, we analyze 45k ad campaigns by 39 parties, both populist and mainstream, in Germany, United Kingdom, Italy, Spain, and Poland.
While populist parties represent just over 20\% of the total expenditure on political ads, they account for 40\% of the total impressions---most of which from Eurosceptic and far-right parties---thus hinting at a competitive advantage for populist parties on Facebook.
We further find that ads posted by populist parties are more likely to reach male audiences, and sometimes much older ones. %
In terms of issues, populist politicians focus on monetary policy, state bureaucracy and reforms, and security, while the focus on EU and Brexit is on par with non-populist, mainstream parties.
However, issue preferences are largely country-specific, thus supporting the view in political science that populism is a ``thin ideology'', that does not have a universal, coherent policy agenda.
This study illustrates the usefulness of publicly available advertising data for monitoring the populist outreach to, and engagement with, millions of potential voters, while outlining the limitations of currently available data.
\end{abstract}

\maketitle \sloppy

\input{intro.tex}

\input{related.tex}

\input{data.tex}
\input{results.tex}

\input{conclusions.tex}

\clearpage
\bibliographystyle{ACM-Reference-Format}
\bibliography{references}

\clearpage
\input{appendix}

\end{document}

%% file: macros.tex
\usepackage{graphicx}
\usepackage[show]{chato-notes}
\usepackage{amsmath}
\usepackage{bbm}
\usepackage{epsfig}
\usepackage{algorithm}
\usepackage{algpseudocode}
\usepackage{url}
\usepackage{hyperref}
\usepackage{tikz}
\usepackage{cleveref}
\usepackage{xspace}
\usepackage[skip=4pt]{caption}
\usepackage{subcaption}
\usepackage{booktabs}
\usepackage{xcolor}
\usepackage{hyphenat} %
\usepackage{siunitx} %
\usepackage{microtype} %
\usepackage{xfrac} %
\usepackage{mathtools} %
\PassOptionsToPackage{capitalise}{cleveref} %
\usepackage{eurosym} %
\usepackage{comment}

\hypersetup{
    bookmarks=true,         %
    unicode=false,          %
    pdftoolbar=true,        %
    pdfmenubar=true,        %
    pdffitwindow=false,     %
    pdfstartview={FitH},    %
    pdfnewwindow=true,      %
    pdftex=true,
    colorlinks=true,
    linkcolor={red!50!black},
    filecolor={green!50!black},
    citecolor={green!50!black},    
    urlcolor={blue!80!black},
}

\newcommand{\spara}[1]{\smallskip\noindent{\bf #1}}

\newcommand{\target}{\textsc{PopuList}\xspace}

\newenvironment{squishlist}
{\begin{list}{$\bullet$}
 {\setlength{\itemsep}{0pt}
     \setlength{\parsep}{3pt}
     \setlength{\topsep}{3pt}
     \setlength{\partopsep}{0pt}
     \setlength{\leftmargin}{1.5em}
     \setlength{\labelwidth}{1em}
     \setlength{\labelsep}{0.5em} } }
{\end{list}}

%% file: intro.tex
\section{Introduction}
\label{sec:intro}

In the past decade, newly-established parties have risen across Europe \cite{guiso2019global,rydgren2018oxford}, with mainstream and centrist parties receiving less attention \cite{hooghe2018cleavage}.
These changes have been attributed to several recent traumatic events, including the Great Recession of 2008 \cite{kriesi2015european}, the migration crisis of 2015 \cite{dinas2019waking}, and the continued economic globalization \cite{colantone2018trade}.
As a result, numerous populist movements have gained popularity by holding the political establishment responsible, and by promoting the sovereignty of ``the people'' (identified, by different movements, as nationality, class, or ethnicity)~\cite{abts2007populism}.
A stark example of such a movement is the 2016 Brexit vote, wherein the United Kingdom voted to leave the European Union. 
The following EU Parliamentary election in 2019 saw high voter participation of about 51\%, and was widely considered a testing ground for the populist movements rising across the continent.\footnote{\url{https://www.politico.eu/article/populist-tide-rises-but-no-flood-eu-elections-2019}}

Beside global transformations and crises, the increasing popularity of populist messages has been linked to the proliferation of social media and online advertising.
Internationally, social media audience concentration has been shown to correlate with the populist vote share~\cite{bennett2021online}, as the profit-maximizing algorithms of the online platforms amplify the most incendiary messages~\cite{esser201728}. %
The filtering mechanisms enabled by social media can favor both the mobilization of crowds of like-minded individuals and the establishment of information sources alternative to mainstream media~\cite{gerbaudo2018social}, thus helping the onset of populist movements.
The leading populist parties, including the Italian 5 Star Movement, the British National Party, and Front National
in France, build websites, maintain social media presence, and often surpass their mainstream counterparts in engagement metrics~\cite{esser201728}.

As the importance of online advertising became apparent, major online platforms took steps to provide increased transparency when it comes to political advertising.
One such effort is the Meta (previously Facebook) Ad Library\footnote{\url{https://www.facebook.com/ads/library/}} which offers a ``comprehensive, searchable collection of all ads currently running from across Meta technologies,'' as well as a historical search for inactive ads around ``issues, elections, or politics''.
Crucially, the dataset captures the activities of the populist political actors around paid content promotion and targeted advertising.
This resource allows us to examine how thematically and demographically cohesive is the populist ``wave'' recorded in the recent decade.

In this paper, we use the Meta Ad Library to examine political advertising around the 2019 EU Parliamentary election in Germany, United Kingdom, Italy, Spain, and Poland---the five largest countries in EU by population (excluding France, where political advertising around elections is forbidden).
In particular, we consider the political parties that have won at least 1 seat or at least 2\% of the votes in a national parliamentary election since 1989, and focus on those identified by \target\footnote{\url{https://popu-list.org}} \cite{rooduijn2019populist} as populist, far-right, far-left, or eurosceptic (note that these labels often overlap).
Our main research questions are the following:
\begin{squishlist}
\item \textbf{RQ1} How are \target parties different from the other ones?
\item \textbf{RQ2} What is common across \target parties in Europe?
\end{squishlist}
We orient our analysis along two main axes: audience demographics and ad content, so we articulate our research in $4$ sub-questions:
\begin{squishlist}
    \item \textbf{RQ1.a}: Are the audiences reached by \target parties different from those reached by other parties? %
    \item \textbf{RQ1.b}: Are the contents of the ad campaigns of the \target parties different than those run by other parties? %
    \item \textbf{RQ2.a}: Are the demographic characteristics of the audience of the \target parties similar across European countries? %
    \item \textbf{RQ2.b}: Are the contents of the ad campaigns of \target parties common across countries? %
\end{squishlist}

We find that populist parties reach a distinctly different audience than other parties, with a higher prevalence of male and older individuals---so much so that it is possible to automatically identify whether an ad is from a populist party judging from the audience it reaches (F1 from $0.64$ to $0.95$). 
Further, the cost to reach this audience is \emph{lower} for populist parties, especially in Germany and Spain.
Confirming previous literature~\cite{noury2020identity,vachudova2021populism}, we find that monetary, state, and bureaucracy reforms feature predominantly in populist messages, as well as security and immigration, which are particularly popular in the eurosceptic and far-right ads; while themes of human rights and environmental issues are underrepresented.
However, commonalities in populist messages across EU are ``thinner'' than the differences between countries: when we take into account countries as confounders, most of the issue focus appears to be country-specific.
In summary, we show an extensive use of Facebook advertising by the populist parties, illustrate its relative cost effectiveness, and demonstrate quantitatively and qualitatively the diversity of populist message across the EU during the election.
We conclude with a discussion of limitations of this data source and point to further opportunities for transparency that will help social media platforms support the democratic process.

%% file: related.tex
\section{Background and related work}
\label{sec:related}

\subsection{Populism}

Populism and anti-elite political stances have existed since the late 19th century, but have become increasingly common at the turn of the 21st century~\cite{albertazzi2008introduction}.
Since the Great Recession of 2008, populism has gained political ground in Europe and elsewhere. 
For instance, in UK, Farage celebrated the Brexit vote as ``a victory for real people''~\cite{williams2016nigel}. %
A new cleavage has emerged in the European parliament: beside left and right, a new separation has developed between populist parties and the mainstream~\cite{noury2020identity}. 
Note, however, that European populism is not necessarily eurosceptic, with notable examples in the Spanish Podemos and the DiEM25 pan-European Movement~\cite{moffitt2018populism}.

\citet{mudde2012populism} have defined populism as an ideology that considers society to be ultimately separated into two groups, the people and the elite; while the elite is often corrupt, for populists, politics should express the ``general will'' of the people.
\citet{laclau2005populism} defines it as an appeal to the entirety of the political community against unresponsive political elites.
Moreover, according to \citeauthor{laclau2005populism}, populists' use of an ``empty signifier'' allows them to join different demands in a single campaign.
Populists often formulate ``the people'' in identities including nationality, class, or ethnicity, and put themselves in opposition to yet other identities~\cite{noury2020identity}.
Given this dichotomous framing, which is moralistic rather than programmatic, populism has been defined a ``thin'' ideology~\cite{mudde2012populism}.
Thus, populism ``is unable to stand alone as a practical political ideology: it lacks the capacity to put forward a wide-ranging and coherent programme for the solution to crucial political questions''~\cite{ernst2019favorable,stanley2008thin}.
As such, populism can be paired with other ideologies~\cite{albertazzi2008introduction}:
\citet{zulianello2020varieties} recognizes radical right populists, who identify the people in ethnic terms; neoliberal populists, who use producerist tones to oppose cultural and bureaucratic elites; left-wing populists, who combine populism with forms of socialism; and ``valence populists'', who focus on issues not positioned on the left-right spectrum.

In Europe, right-wing populism currently represents the most common combination.
From 1979 to 2019, 85\% of the 779 members of populist parties elected to the European Parliament were from the right-wing, compared to 15\% left-wing~\cite{stockemer20202019}.
In our dataset, $55\%$ of Facebook views to populist parties ads are directed to far-right populists, while $19\%$ towards far-left ones.
Radical right populists often have pro-national sovereignty, anti-globalization, anti-immigration stances, with Eastern Europe sometimes also being concerned about the cosmopolitan values coming from the West~\cite{noury2020identity}.
The populist left, conversely, is more concerned about fiscal policy, and takes an anti-austerity stance. 
In combination, populist incumbents may position themselves in the economic left and the cultural right~\cite{vachudova2021populism}.

\subsection{Online political advertising}

Online political advertising has grown significantly over the last years~\cite{sosnovik2021understanding}. 
While many social networks offer online advertising in election campaigns, Meta (previously Facebook) stands out as one of the most influential platforms~\cite{karpf2016analytic}. 
To monitor online sponsored political advertising, these companies have created public collections of ads that run on their platforms. 
In this work, we use Meta Ad Library, a historical collection of politically-relevant advertising that the company surfaces in the aims of greater transparency. %

Political parties post content on Facebook to gain visibility among the electorate on the Internet, and to engage and mobilize their voters online and offline.
For example, \citet{koc2021facebook} %
show that interactivity is important, and responsive party posts on Facebooks are significantly more likely to be shared, liked, and commented on by users.
The audience targeting has also been studied, in the case of anti-immigration advertising in Italy, finding that the political parties promoting anti-immigration messages reach voters similar in age and gender to their voter base~\cite{capozzi2020facebook,capozzi2021clandestino}.

In addition to the public collections created by the companies themselves, there have been efforts by the research community to independently monitor online sponsored political advertising.
\citet{matias2022software} designed a software-supported approach for auditing, which uses coordinated volunteers to analyze political advertising policies enacted by Facebook and Google during the 2018 U.S. election.
A team of volunteers posted auto-generated ads and analyzed the companies' actions, and found systematic errors in how companies enforced policies.
Moreover, an audit of the Ad Library has shown that the platform allows for inaccurate disclosure of advertiser's political advertising activity~\cite{edelson2020security}. 
The authors demonstrate instances of undeclared coordinated activity by ``inauthentic communities'' that are able to fund large-scale advertising campaigns.
Furthermore, concerns have been raised around potentially discriminatory advertising via ``look-alike'' audience matching, which allows the advertiser to define a precise selection of audience members by supplying a list of users with personally identifiable information~\cite{speicher2018potential}.
On the larger scale, political scientists are concerned that the authority to preserve the integrity of democratic deliberation is being ceded to commercial actors ``who may have differing understandings of fundamental democratic norms''~\cite{dommett2019political}.
The role of targeted advertising and differential pricing on reaching a diverse audience has also been criticized for its potential to create political filter bubbles~\cite{ali2021ad,cinelli2021echo}.
Facebook in particular has been under fire for promoting divisive content that is seen as harmful to peaceful demographic processes~\cite{vaidhyanathan2017facebook,entous2017russian}.
For instance, at the beginning of the COVID-19 pandemic, a study found divisive messaging in the Facebook ads around the vaccine and other preventive measures, competing with any potential advertising by public health institutions~\cite{mejova2020covid}.
Further, misinformation was found in the advertising of far-right populist Spanish party VOX around two Spanish general elections in 2019~\cite{cano2021disinformation}.
In this work, we examine the extent of advertising by the populist parties during the European Parliamentary election of 2019, and assess the possible audience targeting, as well as issue ownership, in comparison to other major political parties.

%% file: data.tex
\section{Data}
\label{sec:data}

\subsection{\target}
To better understand the subject of our study, we turn to experts in the field who have compiled a list of parties that have showed populist, extremist, or eurosceptic tendencies.
The \target\footnote{\url{https://popu-list.org}} is the result of close cooperation between academics and journalists, initiated by The Guardian.
The list consists of European parties from 31 countries tagged as populist, far right, far left, and eurosceptic.
Each party can receive multiple tags (e.g., the Lega party in Italy is at the same time tagged as far-right, eurosceptic, and populist).

\subsection{Facebook Ad Library}
We take into consideration all the parties in the list that belong to one of the top-5 countries in EU by population.\footnote{Which includes UK at the time of the elections.}
We exclude France from this list because political advertising is illegal in the country during the six months prior to the elections~\cite{dobber2019regulation}, therefore we do not have enough data for our analysis.\footnote{We only find $3$ ads published by French parties, which were all published after the end of the election on May 26; this fact corroborates our decision to exclude the country.}
The final list of countries is: 
Germany, United Kingdom, Italy, Spain, and Poland.

For each country, we collect all the parties which participated in the 2019 EU elections in May 23-26, and additionally match the same criteria as the \target: at least one seat or at least $2\%$ of votes in a national election.\footnote{Data from \href{https://www.parlgov.org}{ParlGov}.}
The resulting list contains $46$ parties, out of which $17$ are in the \target.\footnote{Here we report the number of single parties, rather than coalitions.}
For each party, we manually identify its main Facebook page and the one of its leader(s).\footnote{In case of coalitions, we take the Facebook page of each of its members and the page of the leader of the coalition.}
This dataset is publicly available\footnote{\url{https://zenodo.org/record/6597765}}.
We find $39$ parties out of $46$ which have a presence on Facebook, for a total of $57$ Facebook pages. 
Among these pages, $21$ belong to parties in the \target ($18$ pages of parties and $3$ of party leaders), and $36$ to parties not in the \target ($27$ pages of parties, $9$ of party leaders).

We use the Meta Ad Library API\footnote{\url{https://www.facebook.com/ads/library}} to retrieve all the ads authored by the pages identified as above, and published between April and May 2019.
Overall, we find \num{44949} ads from the selected five countries (%
\Cref{sec:ap_dataset} shows their volume).
For each ad we retrieve the publication date, the end time of the ad campaign, the text of the ad, the number of impressions (i.e., the number of users who saw the ad, given as a range), the cost of the campaign (again as a range), and the demographic of the audience reached by the ad: gender (male, female, or unknown) and age (in 7 buckets).
Following previous literature~\citep{capozzi2021clandestino}, for values that come in a range (cost and impressions) we take the average of the endpoints of the range, and for open-ended ranges we take the known closed endpoint.

\begin{figure}
    \centering
    \includegraphics[width=\columnwidth]{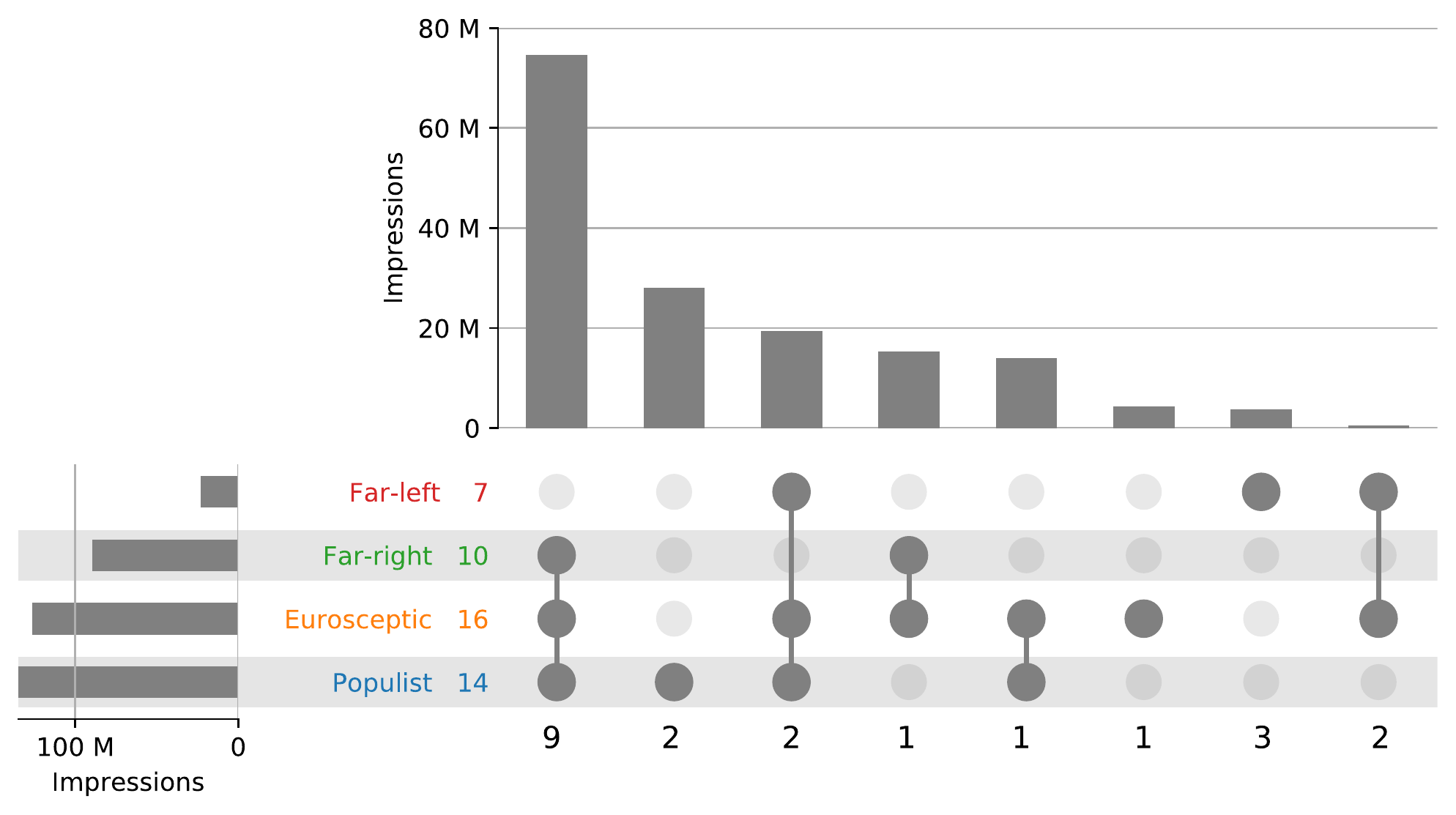}
    \caption{Number of impressions for all the combinations of tags (top bar plot). The lower part of the plot shows the number of Facebook pages for each combination of tag (bottom). For each tag, we show its total number of impressions (left bar plot), and write the total number of pages with that tag (colored numbers). There are 34 pages without any tag, and they obtained a total of almost 242M impressions.}
    \label{fig:tag_combinationsl}
    \Description{Number of impressions for all the combinations of tags (top bar plot). The lower part of the plot shows the number of Facebook pages for each combination of tag (bottom). For each tag, we show its total number of impressions (left bar plot), and write the total number of pages with that tag (colored numbers). There are 34 pages without any tag, and they obtained a total of almost 242M impressions.}
\end{figure}

By combining these two data sources (PopuList and Facebook ads), we can visualize the most frequent combinations of tags present in political advertisement in \Cref{fig:tag_combinationsl}.
The most common tag is eurosceptic (16) followed by populist (14), however the latter accrues a larger number of impressions overall.
Far-left parties are neither common nor very prominent in terms of impressions.
The most common combination of tags is the triplet populist, eurosceptic, far-right (9), while we have only 2 representatives of the combination with complementary ideology (populist, eurosceptic, far-left).
Parties tagged with the far-right combination obtain approximately 3 times more impressions that the ones with the far-left one.
\Cref{tab:ad_stats} reports more detailed statistics about the dataset.

In order to fairly quantify the impact of each ad and party, in the following, rather than looking at individual parties or Facebook pages, we take as the unit of analysis a single ad impression.
That is, when aggregating ads by parties with the same tag, their weight is proportional to their reach.
For each ad, we then focus on their main characteristics available: audience and content.

\subsection{Content annotation}
\label{sec:content-annotation}
To characterize the most important issues mentioned in the collected ads---both those from parties in PopuList, and those not associated with these characteristics---we use inductive qualitative coding (for an overview of the process, see the introduction by \citet{linneberg2019coding}). 
To overcome linguistic barriers, we use Google Translate to obtain an English translation of each ad, available during annotation beside the original version.
We remove near-duplicate ads (i.e., based on the similarity of lemmatized ad text).
Then, we use an iterative, collaborative process to create the codebook, wherein an initial set of codes from the literature~\cite{mudde2017populism, rooduijn2019populist} is augmented and iteratively refined by all 5 authors.
Specifically, we begin the process with a sample of ads from all countries and parties, and qualitatively explore them by identifying major codes and more detailed codes for issues either explicitly or implicitly mentioned in each ad (there could be multiple). 
In this effort, we are guided by the previous literature on the common emphases of populist movements, which have been shown to include national sovereignty, immigration, and cultural values on the right, and fiscal policy and austerity on the left~\cite{noury2020identity}.
For example, a major code ``human rights'' may have several sub-codes including ``women's rights'', ``freedom of speech'', and ``voting rights''. 
Then, we label all the considered ads following open coding procedures~\cite{linneberg2019coding}, with periodical discussions to cluster similar concepts and reach a consensus. 
Our codes are determined by all annotators, first separately, and merged together upon discussion; as such, no annotator agreement can be computed.
This way, we annotate all ads from PopuList parties, totaling \num{1898} ads; of those, \num{337} ads are related to local rather than European elections, and are thus removed.
We publicly release this labelled data set.\footnote{\url{https://zenodo.org/record/7594103}} \enlargethispage{0.3\baselineskip}
\Cref{sec:ap_codes} reports the final list of top-level and sub-codes.
These codes reflect issues discussed not just by parties in \target, but all others in the selected countries.

Finally, the remaining ads are annotated with this coding scheme, with unclear examples discussed together (because of the open nature of the task, no annotator agreement was computed).

%% file: results.tex
\section{Results}
\label{sec:results}

\subsection{Characterizing \target parties}

To answer \textbf{RQ1.a}, we characterize the parties according to the gender and age of the audience reached, and additionally by looking at the cost per impression of the ads, which is a proxy for the value of the audience to the given party.

\begin{figure*}[t]
    \centering
    \includegraphics[width=0.75\textwidth]{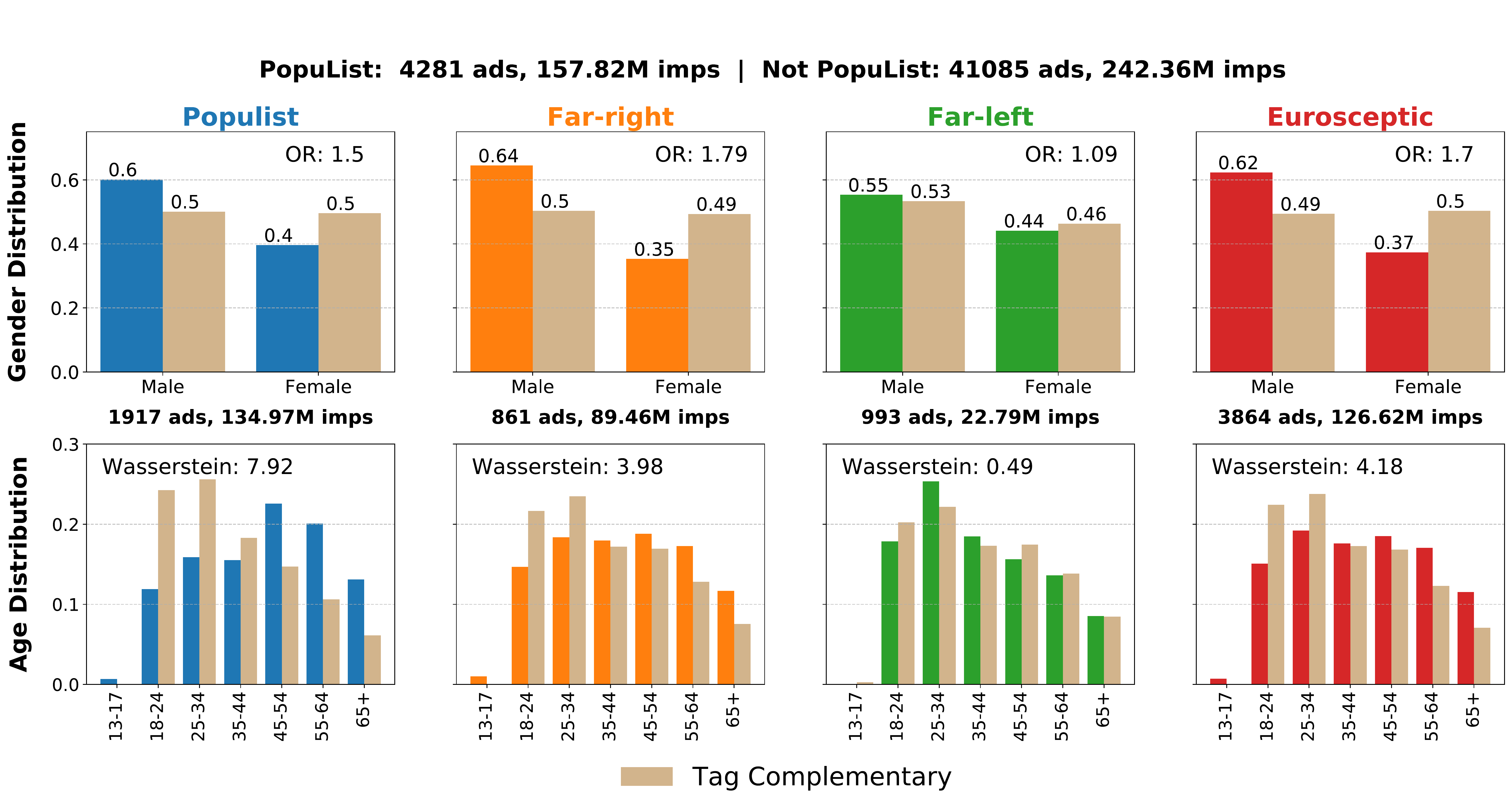}
    \caption{Comparison between the demographic reach by age and gender for each tag and all ads without a given tag, aggregated over all countries.
    For gender, we measure the male-to-female odds ratios.
    For age, we use Wasserstein distances.}
    \label{fig:demog_oddsratios}
    \Description{Comparison between the demographic reach by age and gender for each tag and all ads without a given tag, aggregated over all countries.
    For gender, we measure the male-to-female odds ratios. For age, we compare the distribution by using Wasserstein distance.}
\end{figure*}

\begin{figure}
    \centering
    \includegraphics[width=0.5\columnwidth]{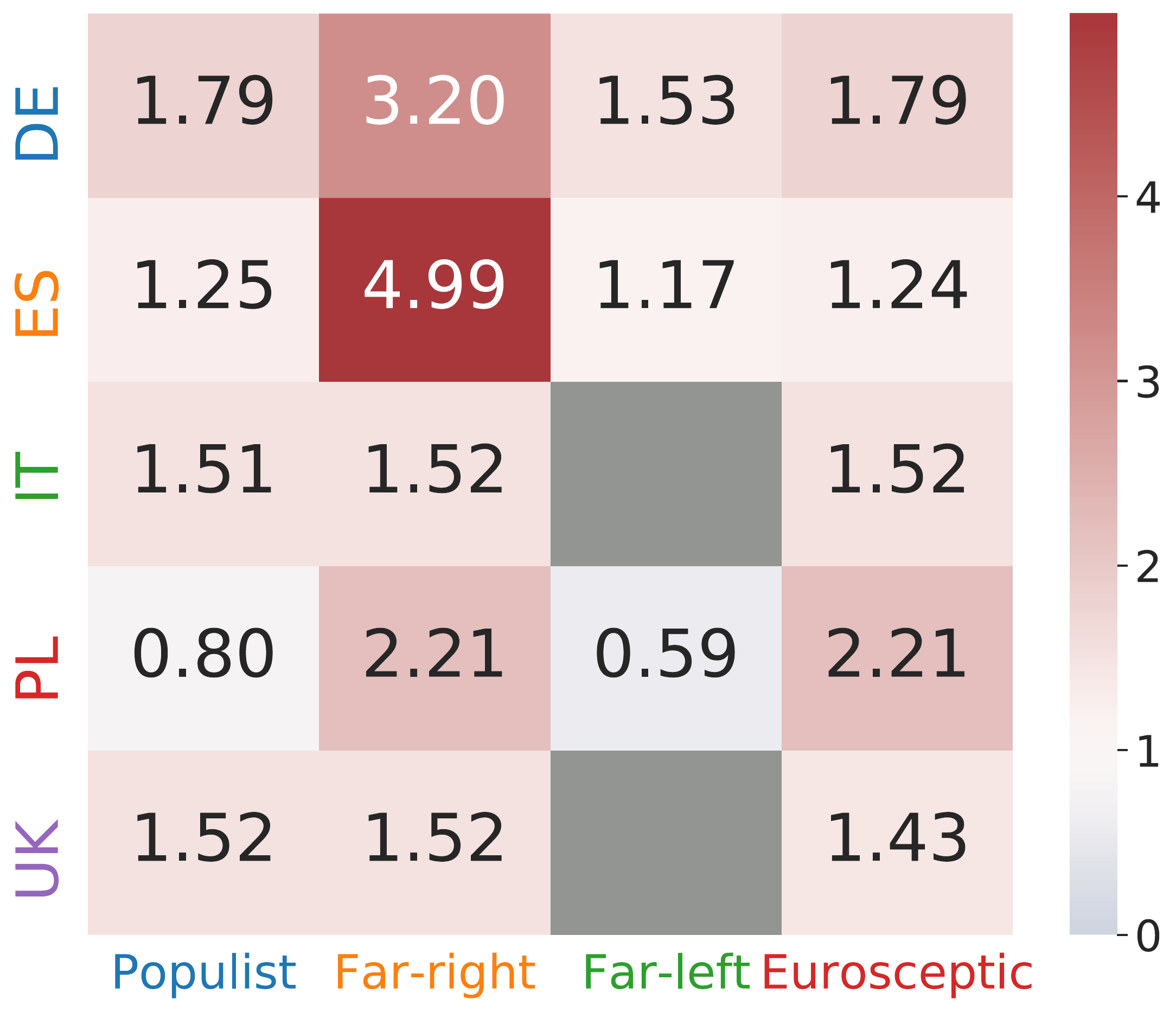}
    \caption{Country-specific gender odds ratio (male to female) between each tag and its complementary.}
    \label{fig:demographic-countries}
    \label{fig:wass_odds}
    \Description{Country-specific gender odds ratio (male to female) between each tag and its complementary}
\end{figure}

\begin{figure}
    \centering
    \includegraphics[width=0.95\columnwidth]{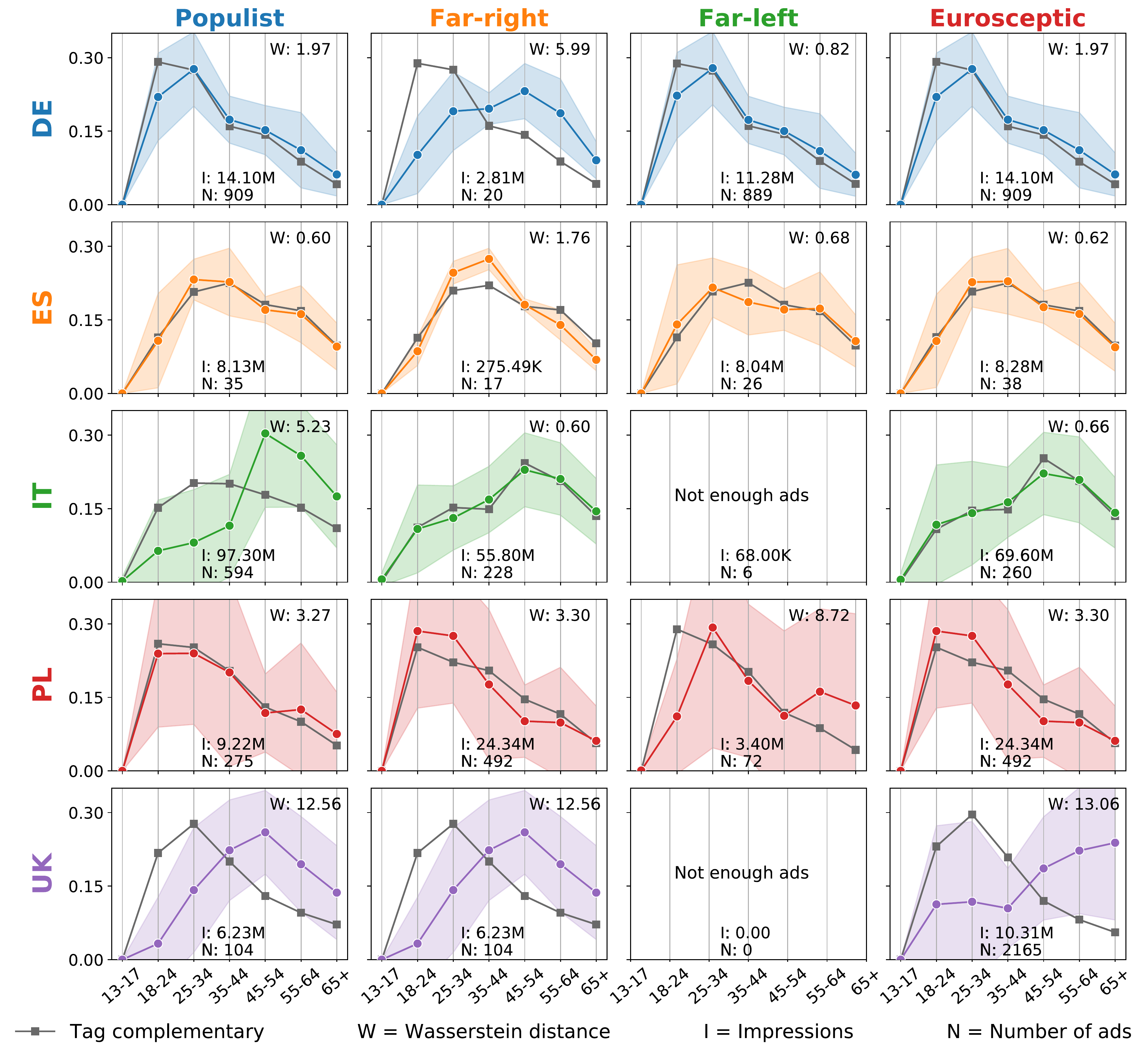}
    \caption{Country-specific age distributions (mean and SD) between each tag and its complementary.}
    \label{fig:age_distribution}
    \Description{Country-specific age distributions (mean and SD) between each tag and its complementary.}
\end{figure}

\subsubsection{Demographics}
We compare the audience reached by \target parties to the one reached by other parties.
\Cref{fig:demog_oddsratios} shows the demographic reach for each tag, compared to the one of all ads without the given tag (called tag complementary), aggregated over all countries.
As summary statistics, for gender, we use odds ratios to measure the increase in likelihood of a viewer of the ad being male.
For age, we use the Wasserstein distance between the distributions, which represents the difference (in years) between the average audience of \target parties compared to the others.

Overall, populist, far-right, and eurosceptic parties tend to reach more male audiences ($+50\%$, $+79\%$, $+70\%$ respectively).
In addition, the age Wasserstein distance values of $7.92$, $3.98$, and $4.18$, respectively, indicates that these parties reach older audiences on average.
These similar findings could be partially explained by the high overlap between these three tags, as shown in \Cref{fig:tag_combinationsl}.
Conversely, far-left parties reach a balanced audience in terms of gender ($+9\%$) and similar ones in terms of age (Wasserstein distance of $0.49$).
The small difference we observe is in the opposite direction: far-left parties reach a larger fraction of people of age 25-34 across Europe.

\spara{Country-specific demographics.}
To assess the variability of such demographic characteristics across Europe, we replicate the analysis by country.
For far-left in Italy and UK we do not have enough ads to run the analysis (fewer than $10$ ads), therefore we omit the results for these two cells. %
Looking at gender distribution, \Cref{fig:wass_odds} shows, for each country, the male-to-female odds ratio in the audience of each tag.
Beside confirming the result from the previous analysis---most \target parties generally reach an audience with a higher presence of males---we observe that in particular far-right parties, especially in Germany and Spain, have a markedly more male audience.
The exception to this pattern is Poland, the only country where a tag presents a larger female audience, in the case of both populist and far-left parties (note that these two tags correspond to distinct sets of parties in this country).

Regarding age, \Cref{fig:age_distribution} shows, for each country, the normalized age distribution of the audience reached by parties with each tag, compared to all the parties without the given tag.
The shaded area represents the standard deviation of the reached audience for each bucket, computed across ads.
We see regional differences within the same tag.
For instance, while eurosceptic parties in Poland reach a younger audience, the opposite is true in UK.
A similar pattern can be seen for far-right between Poland and Germany.
UK shows the largest differences in audience age between the parties, with \target ones generally reaching an older audience.
Overall, no consistent cross-country pattern can be identified.

To understand if these differences are significant when taken together, we build a machine learning model that distinguishes whether the ads comes from a party with a specific \target tag given its reached audience. 
In general, the demographic-based classifier manages to distinguish populist, far-right, and eurosceptic ads with a sufficient level of accuracy (F1 $\geq 0.64$).
However, far-left ads are less distinguishable than their counterparts.
See \Cref{sec:ap_model_1} for country-specific classification results.

\begin{figure}
    \centering
    \begin{subfigure}[b]{\columnwidth}
        \includegraphics[width=\columnwidth]{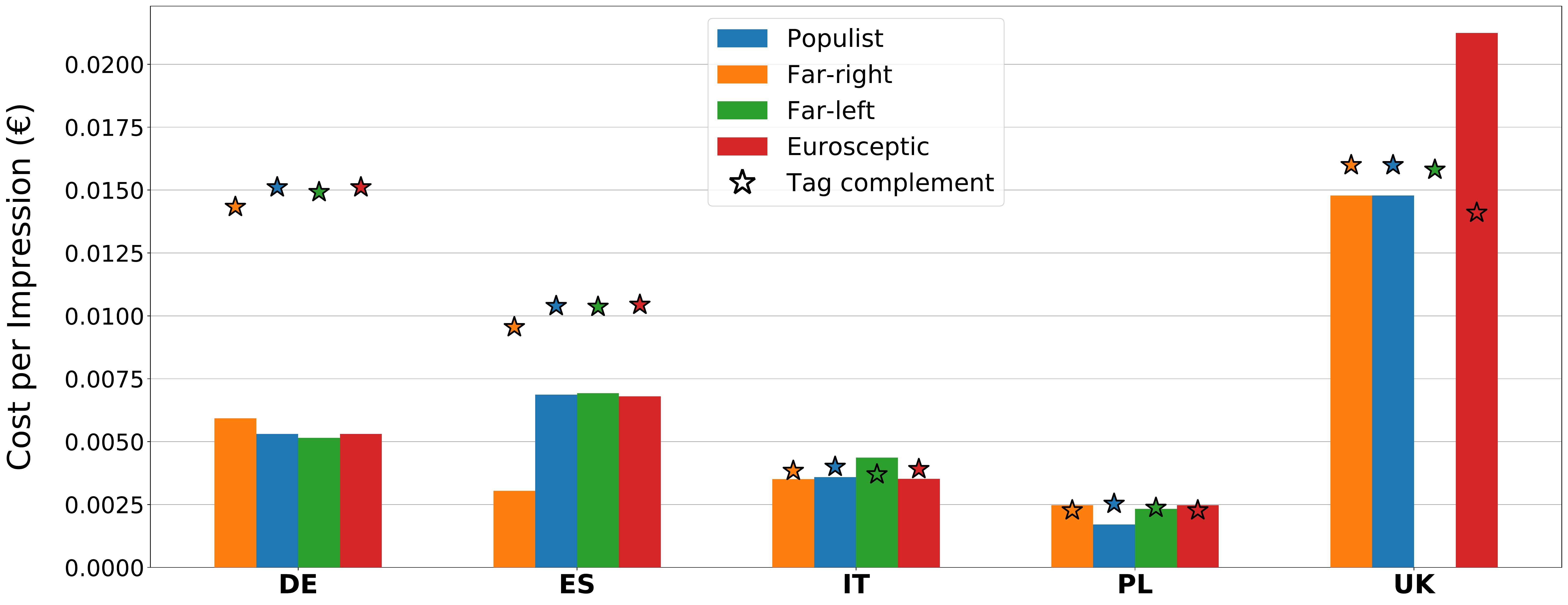}
        \vspace{-18pt}
        \caption{}
        \vspace{5pt}
        \label{fig:cost_per_impression}
    \end{subfigure}
    \begin{subfigure}[b]{\columnwidth}
        \includegraphics[width=\columnwidth]{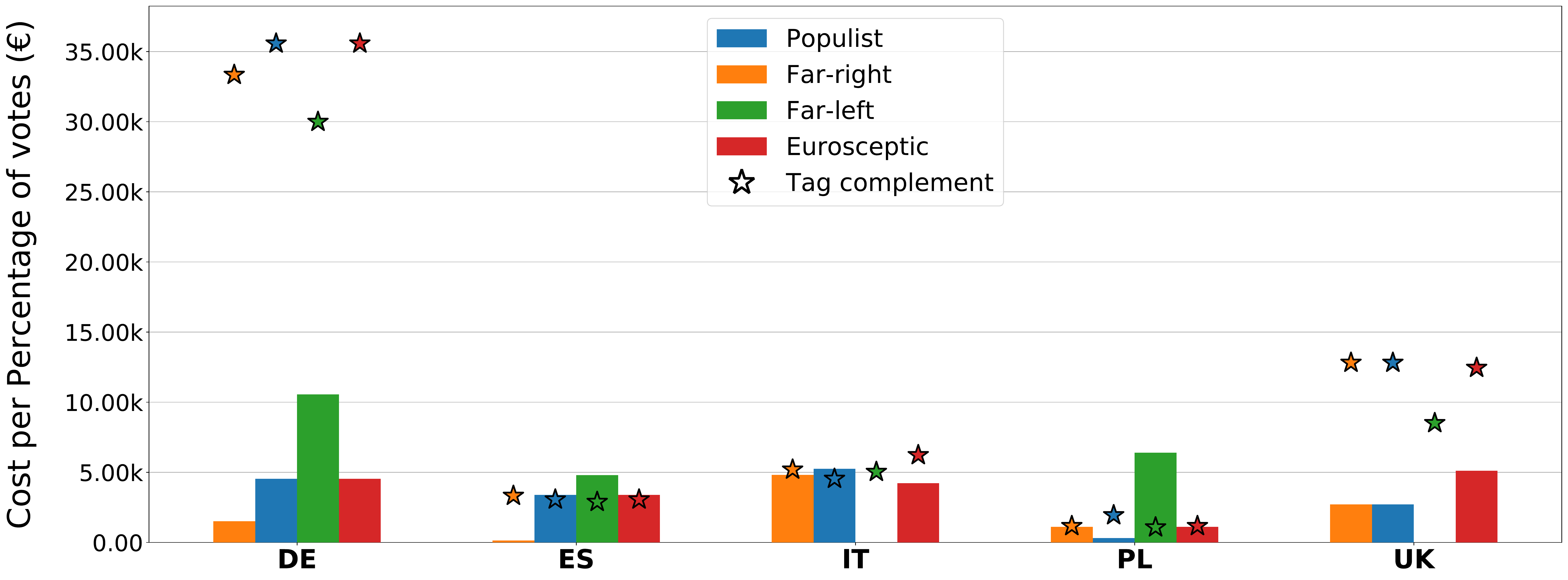}
        \vspace{-18pt}
        \caption{}
        \label{fig:cost_per_vote}
    \end{subfigure}
    \caption{Cost~(\subref{fig:cost_per_impression}) per impression and~(\subref{fig:cost_per_vote}) per percentage of vote for each tag for each country. Star mark represents the tag complementary.}
    \label{fig:cost_per_impression_vote}
    \Description{Cost per impression and per percentage of vote for each tag for each country. Star mark represents the tag complementary.}
\end{figure}

\spara{Advertisement effectiveness.}
\Cref{fig:cost_per_impression} shows the cost per impression for ads with different tags and countries, compared to their complementary ones (marked with a star).\footnote{All the costs are converted to Euros by using the average exchange rate in May 2019.}
In both Germany and Spain, \target ads are noticeably cheaper than their complement (especially for far-right in Spain).
Conversely, there is no difference for Italy, Poland, and most of the tags in UK.
Interestingly, eurosceptic ads in UK are the only ones that are more expensive than their counterpart.
Note that we are comparing to other political ads, so the price difference can be attributed to some characteristic of the ads (either how they are targeting their audience, their content, or timing).
When looking at the effectiveness of advertising at the ballot (\Cref{fig:cost_per_vote}), in Germany, and to some extent in UK, non-\target parties spent more per vote than \target parties.
The original data shows that this pattern is driven by establishment parties (CSU-CDU, SPD, and FDP) who spent much more than others on social media, and reaped limited benefits at the vote.
A similar pattern is present in UK with LibDem and Tories out-spending other parties.
In Spain, instead, while the cost per impression is lower for \target parties, this advantage does not translate into an improved outcome in votes.
Conversely, in UK, all parties show similar cost per impression, but \target parties fared better in elections.
Clearly our analysis is not causal as it ignores hidden confounders (including actual persuasiveness of online ads~\cite{kruikemeier2016political}), however it is indicative of potential differences across countries.

Overall, in 2 out of 5 countries (Germany and Spain), \target ads are cheaper, with a single example for which the opposite holds (eurosceptic in UK).
The impressions for ads on Facebook are split $40/60$\% between \target and non-\target parties, respectively.
This proportion mirrors the results of the 2019 EU elections, where \target parties obtained roughly $44\%$ of the votes (see \Cref{tab:total-popu-list})---even though no direct causal link can be claimed.
However, contrary to the common narrative, non-\target parties have spent \emph{more} on advertising, mainly driven by large expenses of establishment German parties.
Finally, there are large regional differences in the costs of advertising, which however do not seem immediately explainable by the economic development of the countries.

\subsubsection{Content}
We now turn our attention to content in order to answer \textbf{RQ1.b}: are the contents of the ad campaigns of the \target parties different from those run by other parties?
To do so, we use the issues and sub-issues we have identified through content annotation (\Cref{sec:content-annotation}).
The issues of Security and Immigration are dominated by parties that are at the same time eurosceptic, far-right, and populists.
The issue of Institutions (e.g., state reforms) is owned by populist parties in general.
Far-left parties, instead, tend to focus on Economy, Human Rights, and Environmental issues.
Environment is underrepresented in far-right parties, while issues related to the European Union are underrepresented in far-left parties.
See \Cref{sec:ap_content} for a visualization.

\begin{figure*}[t]
    \centering
    \includegraphics[width=\textwidth]{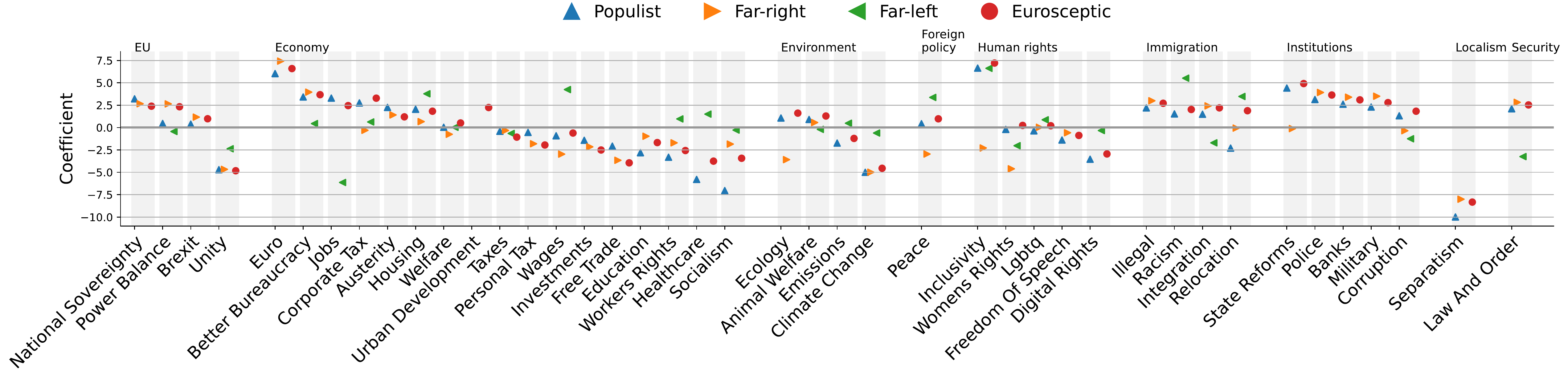}
    \caption{Coefficients for each sub-issue in a logistic regression model aimed at identifying whether the author of an ad is a \target party. Each \target tag has a separate model, represented by different symbols and colors. Coefficients are sorted grouped by issue and sorted decreasingly by `populist' weight within the group. Ads are weighted by number of impressions.}
    \label{fig:regressions-1}
    \Description{Coefficients for each sub-issue in a logistic regression model aimed at identifying whether the author of an ad is a PopuList party. Each PopuList tag has a separate model, represented by different symbols and colors. Coefficients are sorted grouped by issue and sorted decreasingly by `populist' weight within the group. Ads are weighted by number of impressions.}
\end{figure*}

To identify and quantify which specific sub-issues are particularly associated with \target parties, we use a simple regression model.
Specifically, based on the sub-issues determined for each ad, we build a logistic regression model aimed at predicting whether an ad is authored by a party with a given \target tag.
Since we wish to model the average case for a European Facebook user, we again weight each tag by its number of impressions.
For this research question we are not interested in country-specific effects, but rather whether \target parties own specific issues across all considered countries.
Therefore, we ignore country-specific effects in this model; we investigate those in \Cref{sec:rq2b}.

\Cref{tab:auc} reports the goodness of fit obtained by this model for each of the \target tags, measured as the area under the ROC curve (rightmost column).
The values range from $0.67$ to $0.80$, which suggest that the topic of an ad can often be enough to distinguish a \target ad from non-\target ones, regardless of the country of origin.
This correlation is stronger for far-right and far-left parties, that seem to share more affinity across European countries---which is to be expected, considering they are both more well-established and historically well-defined than populism.

\Cref{fig:regressions-1} shows the coefficients for each sub-issue in these models.
Such coefficients quantify how much more likely an ad is to belong to a given tag given its sub-issues.
More abstractly, they indicate which topics are more important for \target parties than other parties, across all the considered countries.
By looking at these coefficients, we can sketch what platform \target parties were promoting through Facebook ads in the elections.
First, far-right, eurosceptic, and populist parties are very often grouped together (remember that parties with all three tags represent a larger share than parties with only one, \Cref{fig:tag_combinationsl}).
In general, these parties focus more on the Euro, bureaucracy, illegal immigration, law \& order, and institutions such as police and the military;
while, they advertise significantly less on climate change and regional separatism.

On the topics related to the EU, these parties talk more about problems related to the institutional structure of the Union: power balance, Brexit, and national sovereignty.
Moreover, all \target parties (i.e., including far-left) advertise less than other parties on the topic of European Unity.
Focusing specifically on populist parties, their typical issues ads are job creation, corporate taxes, austerity, and institutional reforms.
On the contrary, they tend to ignore issues such as healthcare, workers' rights, and education.

Finally, far-left parties behave differently than other \target parties.
In particular, the most identifying far-left issues are economic (housing, wages, healthcare, workers' rights) and related to human rights (peace, inclusivity, racism).
They also focus less on law \& order and job creation (job creation rhetoric has often been attacked from the left as a dog whistle for the wealthy class~\cite{peck2014you}).

\begin{table}[t]
    \caption{Fraction of total estimated expenditure, impressions, and votes, for \target parties.}
    \label{tab:total-popu-list}
    \begin{tabular}{lll}
    \toprule
    {} &               \target &                  Others \\
    \midrule
    Est. expenditure (€) &     \num{771776} (21\%) &    \num{2879761} (79\%) \\
    Est. impressions     &  \num{160662868} (40\%) &  \num{239416693} (60\%) \\
    Votes                &   \num{44448658} (44\%) &   \num{56629116} (56\%) \\
    \bottomrule
    \end{tabular}
    \vspace{-\baselineskip}
\end{table}

\begin{table}[t]
    \caption{Model fit (in-sample AUC ROC score) for the different regression models tested.}
    \label{tab:auc}
    \centering
    \begin{tabular}{lrrr}
    \toprule
\target & Issue + country & Issue + country & Sub-issue +  \\
tag &  random slopes & rand. intercepts & no country \\
    \midrule
Populist & $\mathbf{0.879}$ & 0.833 & 0.671 \\
Far-right & $\mathbf{0.887}$ & 0.864 & 0.767 \\
Far-left & $\mathbf{0.869}$ & 0.848 & 0.802 \\
Eurosceptic & $\mathbf{0.828}$ & 0.772 & 0.683 \\
    \bottomrule
    \end{tabular}
    \vspace{-\baselineskip}
\end{table}

\subsection{Similarities of \target across Europe}
Finally, we evaluate the similarity of Facebook ads from \target parties across countries.
By estimating common traits, we also weigh how prominent they are with respect to cross-country differences.
We do so in terms of demography reached (\Cref{sec:cross-country-demog}, \textbf{RQ2.a}), and then in terms of ad content (\Cref{sec:rq2b}, \textbf{RQ2.b}).

\subsubsection{Cross-Country Demographic Similarity}
\label{sec:cross-country-demog}
To evaluate similarities across the audiences of ads from each \target tag in different countries (\textbf{RQ2.a}), we compare their classification models.
For each country and \target tag, we train an SVM classifier to distinguish whether an ad from country $X$ is run by a \target-tag party (e.g., populist), based on the demographic of its audience (we choose SVM, as it had a better performance over Logistic-Regression at 78\% F1 vs.~67\%).
Then, we compare the coefficients obtained by such models: if the two classifiers attribute similar weights to each age and gender group, it means their ads demographics are similar.

Across countries, a few interesting patterns emerge.
UK populist, far-right, and eurosceptic ads shows a marked similarity in their audiences to German far-right ones (AfD).
Moreover, populist parties in Poland also target the same demographic as Spanish far-left: this effect is explained by a similar male-to-female ratio as both reach a more female audience (\Cref{fig:demographic-countries}).
Finally, Italian populist and eurosceptic ads exhibit some similarity with all Spanish ads.
\Cref{sec:ap_demo_similarity} reports a visualization of the cosine similarity of the SVM coefficients of each classifier across countries and tags.
We further test this hypothesis by formulating a \emph{domain adaptation} classification task.
We obtain low performance scores, which suggests that the differences across European countries are more prominent than any similarities in their demographics (see \Cref{sec:ap_demo_similarity}).

\begin{figure*}
    \centering
    \includegraphics[width=0.9\textwidth]{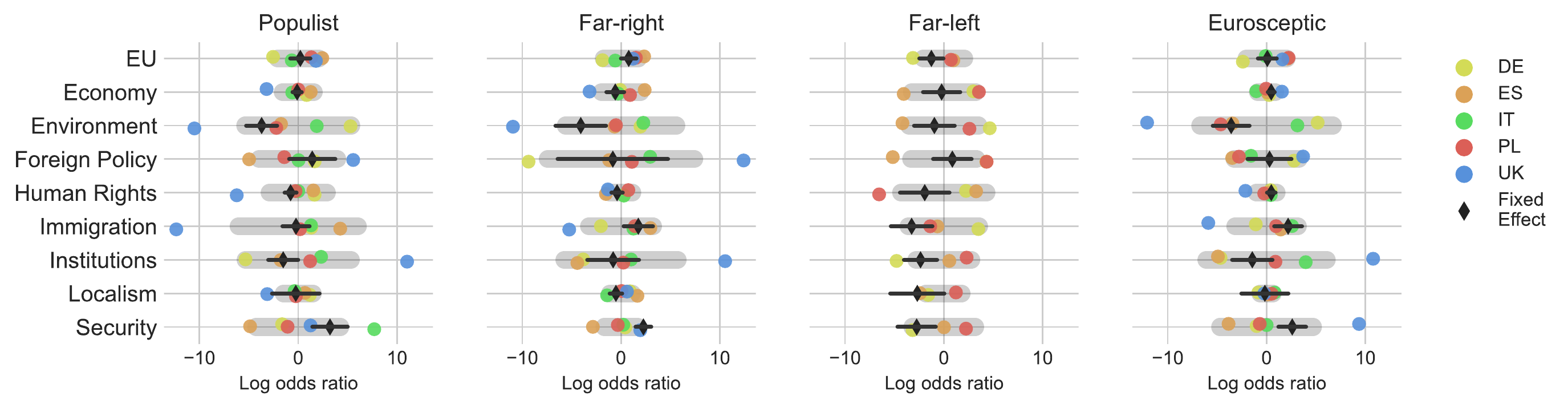}
    \caption{
        Coefficients of logistic regression model with random slopes for each country and issue combination (colored dots), and fixed effect for each issue (black diamond).
        Solid black line represents the standard error of the fixed effects.
        The shaded gray area represents the standard deviation of the random slopes: larger areas indicate larger differences among countries.
    }
    \label{fig:randslopes}
    \Description{Coefficients of logistic regression model with random slopes for each country and issue combination (colored dots), and fixed effect for each issue (black diamond).
        Solid black line represents the standard error of the fixed effects.
        The shaded gray area represents the standard deviation of the random slopes: larger areas indicate larger differences among countries.}
\end{figure*}

\subsubsection{Cross-country content similarity}
\label{sec:rq2b}
We finally turn our attention to \textbf{RQ2.b}: are the contents of the ad campaigns between all of the populist parties common across countries?
To answer this question, similarly to the previous section, we model the problem as a logistic regression where we predict whether an ad is of a given tag (e.g., populist) given the issues it talks about.
This way, the coefficients of the model tell us which issues are important for \target parties across all the considered countries---i.e., whether viewing an ad about a certain topic makes it more likely to be looking at an ad of a given tag.
In particular, we employ a generalized linear mixed effects model (GLMM) to control for country-specific effects.
The specification of the model has fixed effects for each issue, and \emph{both} random intercepts and slopes for each combination of country and issue.\footnote{As before, we weight ads by their number of impressions, and use a Bayesian method to infer the parameters of model, so to add Gaussian priors for the distributions of the random effects, via the \href{https://rdrr.io/cran/blme}{blme} package.}
Compared to a simpler model with only random intercepts, this version shows a better fit in terms of AUC (see \Cref{tab:auc}, first and second columns), and is also statistically significantly better according to a likelihood ratio test ($p<10^{-6}$).\footnote{We were unable to estimate a similar model that uses sub-issues as independent variables, due to the high granularity of the sub-issues which requires a very large amounts of parameters to be estimated.}
We give a qualitative interpretation of the model next.

The fixed effects (FE) coefficients can be interpreted as a standard logistic regression.
The random intercepts simply account for the different baseline prevalence of each tag in each country.
For the random slopes, instead, the assumption is that talking about an issue has a different effect in each country.
That is, we assume that the effect of the country is also mediated through differential effects on the issues.
For example, the issue of institutional reforms is a main talking point for populist, far-right, and eurosceptic parties in UK, such as the Brexit party, but not as much in other countries.\footnote{See for example \url{https://www.facebook.com/ads/library/?id=451295515626544}}
\Cref{fig:randslopes} shows the coefficients for this model, with fixed effects in black (the solid line represents the standard error), and the posterior estimate the random intercept for each country in a different color.
The shaded gray area represents the standard deviation of the random slopes (the main parameter estimated by the model): larger areas represent a more marked difference across countries.

We first focus on the fixed effects, which represent the commonalities across Europe as identified by this model.
There are just a few cross-country effects consistently shared by \target parties.
Populist, far-right, and eurosceptic parties consistently under-represent environmental issues in their ads, while over-representing those related to security.
On this latter issue, far-left parties behave in a diametrically opposite way, and consistently neglect security-related issues.
Finally, immigration is the focus of far-right and eurosceptic parties across Europe, while not a topic of interest for far-left parties.
Other effects are either small or not significant.

Let us now compare the standard deviation of the random effects, represented by the shaded gray area, across issues and tags.
Several areas are quite large, and their effect is comparable to or larger than the corresponding fixed effect.
This result indicates that there is a large difference among countries on these issues.
In particular, environment, foreign policy, immigration, institutions, and security present rather dispersed random effects.
Conversely, for the other issues the spread is lower, but these also correspond mostly to small or non-significant fixed effects.
Overall, the picture painted by the model is that the differences across countries are stronger than the similarities among them.

%% file: conclusions.tex
\section{Conclusions and Future Work}
\label{sec:conclusions}

The findings of this study confirm the ephemeral, or ``thin'' nature of populism, as postulated by political science literature~\cite{ernst2019favorable,stanley2008thin}.
Although we find some commonalities among \target parties, especially within the far-left and far-right, the country-level differences seem to be more important.  
Intuitively, since populism positions itself as anti-establishment, it follows that there would be little coordination between the parties across national borders in the form of a supranational entity.
These findings further support the observation that much of European populist rhetoric is nativist or nationalist in general~\cite{moffitt2018populism}, although it was challenging to extrapolate this stance reliably from the advertising material labeled in this study.
Still, we find a great overlap between the populist, eurosceptic, and far-right labels in the data, which indicates the stances associated with the rejection of the pan-European political vision are heavily entwined with the conservative worldview.

Secondly, this study confirms the extensive use of social media advertising by the \target parties.
Compared to other ones (see \Cref{fig:sankey}), \target parties advertised extensively on Facebook, being responsible for about the $40\%$ of the impressions, while achieving a $44\%$ share of votes in the European Parliament, with a much smaller expenditure ($21\%$ of the total budget).
The lower cost to reach their audience for some of these parties (mainly those in Germany and Spain) is the main reason behind this result.
One possible cause is that the \target parties in Germany and Spain might be targeting audiences that are cheaper to reach.
In fact, it has been shown that female audiences have a higher click-through rate, and are therefore more expensive at auction time~\cite{ali2019discrimination}.
Also, a low budget constraint on a given ad skews the audience towards males because of the aforementioned effect~\cite{ali2019discrimination}.
Another possible cause is that Facebook considers these ads as more engaging for the audience, and therefore serves them at a lower cost~\cite{ali2021ad}; also in this case, small ad budgets exacerbate the effect.

We expect advertisement costs to relate to economic indicators in each country, however, this is only partly true.
Ads in Poland are the cheapest, consistent with it being the only developing country in our sample (according to the IMF).
However, e.g., Germany had a higher GDP per capita than UK in 2019, and Italy a higher one than Spain.
Additional information about the target audience specification is needed to understand the reason for this drastic spread in cost per impression.
It may be that it is easier to reach the audiences which interest \target parties, or that there are few other advertisers interested in these audience, which pushes the price down. %
Notably, the cost per impression is much higher for the UK Eurosceptic parties, which suggests the conditions of the targeting in post-Brexit debate may differ from the other settings.

This study has several limitations.
First, the definition of populism is debatable.
We use labels provided by The \target, a collaboration between social scientists and journalists, however, populist messages may be expressed by politicians across the board.
Second, the selection of five countries limits the generalizability of the results presented here.
Considering more countries may show a stronger party label effect on issue ownership, with a weaker country-specific effect.
The study also pertains to a peculiar time: it is possible the messaging changes when no immediate voting events are upcoming, and instead focuses on growing the supporter base.
Third, insights presented here are somewhat limited by the information Meta publishes in its Ad Library.
We are not certain about the methodology used to include ads in the library, with the possible exclusion of important issues and political players. 
For instance, there are only 20 ads provided for Germany's AfD party, despite news coverage of a more extensive use of the platform~\cite{holroyd}. %
It is possible the party uses other ways to communicate, such as through organic posts.
Further, we already mentioned the lack of information on the audience targeting, some of which may concern the interests, family situation, immigration status, and other personal characteristics (the ad platform has been used to track international migration~\cite{zagheni2017leveraging} and health-related behaviors~\cite{araujo2017using}). 

Finally, we hesitate to make prescriptive statements around potential censorship of political advertising by the platforms or governments. Instead, data and algorithmic transparency may empower people to gauge why they are presented with certain messages.

%% file: appendix.tex
\section*{Appendix}
\appendix
\label{sec:si}

\setcounter{figure}{0}
\setcounter{table}{0}    
\renewcommand{\thefigure}{A\arabic{figure}}
\renewcommand{\thetable}{A\arabic{table}}

\section{Terminology}
\label{sec:ap_terminology}

Here, we report a description of the key terms used in the paper.
\begin{itemize}
    \item Ad cost: For each ad campaign, Facebook Ads Library API returns a range of the cost paid by the author. For each ad we compute the cost as $range\_min + (range\_max - range\_min)/2$. The cost per each impression is determined by a bidding process internal to Facebook, up to exhaustion of the ad campaign budget. \textit{Cost per percentage votes} and \textit{cost per impression} in figure \ref{fig:cost_per_impression_vote} are computed as $(total\_ad\_cost)/(percentage\_of\_votes\_in\_election)$ and $(total\_ad\_cost)/(total\_impressions)$ respectively.
    \item Ad impression: Impressions measure how often an ad was on screen for the target audience. Facebook Ads Library API returns a range of impressions received by the ad. For each ad we compute the number of impressions as $range\_min + (range\_max - range\_min)/2$. Impressions may reach a user multiple times, so the number of impressions does not equal the number of unique users reached.
    \item Tag: with the term tag, we refer to one of the four tags (populist, far right, far left, and eurosceptic) proposed by the \target\footnote{\url{https://popu-list.org}} project described in section \ref{sec:data}.
\end{itemize}

\section{Dataset Statistics}
\label{sec:ap_dataset}

\begin{figure}[H]
    \centering
    \includegraphics[width=\columnwidth]{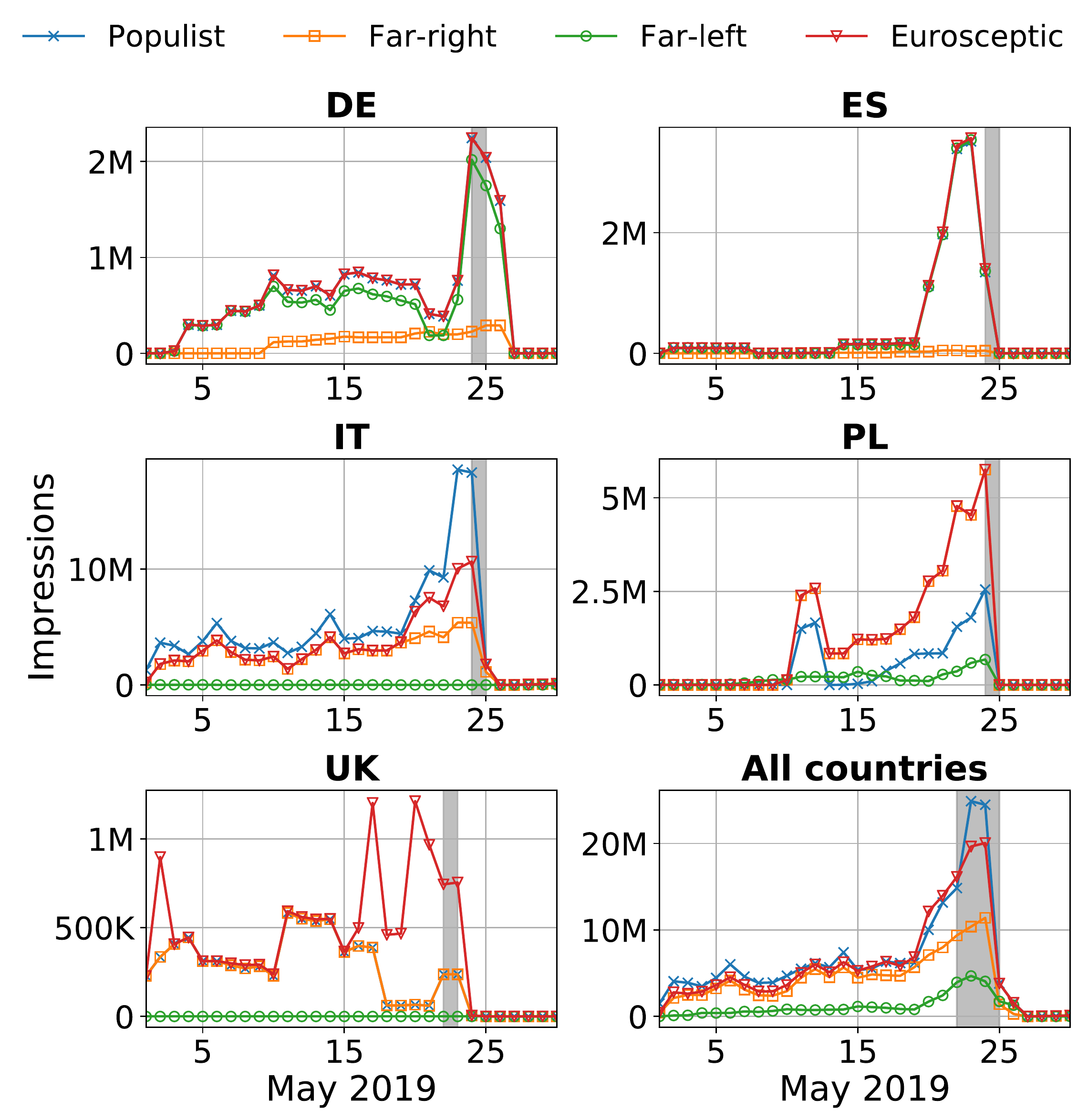}
    \caption{Impressions over time of tags for each country. Election dates are highlighted in gray.}
    \label{fig:volumes}
    \Description{Impressions over time of tags for each country. Election dates are highlighted in gray.}
    \vspace{-\baselineskip}
\end{figure}

Here, we report some statistics about the data set.
\Cref{fig:volumes} shows the temporal evolution of the ad impressions by given tag.
\Cref{tab:ad_stats}, instead, shows the raw numbers of ads and impressions per tag.

\begin{table}[H]
\caption{Total number of ads (and impressions) by country and tag.
The `All' column shows the total number of ads collected, including ads from non-\target parties.}
\footnotesize
\centering
\label{tab:ad_stats}
\begin{tabular}{@{}l@{\quad}r@{\quad}r@{\quad}r@{\quad}r@{\quad}r@{\quad\quad}r@{}}
\toprule
Country &     Populist &     Far-right &      Far-left &   Eurosceptic &  \target &    All \\
\midrule
DE    &  909 (14.1M) &    20 (2.8M) &  889 (11.3M) &   909 (14.1M) &        909 &  35348 \\
ES    &    35 (8.1M) &  17 (275.5k) &    26 (8M) &     38 (8M) &         43 &    514 \\
IT    &  594 (97.3M) &  228 (55.8M) &    6 (68k) &   260 (69.6M) &        600 &    893 \\
PL    &   275 (9.2M) &  492 (24.3M) &    72 (3.4M) &   492 (24.3M) &        564 &    773 \\
UK    &   104 (6.2M) &   104 (6.2M) &        0 (0) &  2165 (10.4M) &       2165 &   7418 \\[2pt]
All   &         1917 &          861 &          993 &          3864 &       4281 &  44946 \\
\bottomrule
\end{tabular}
\end{table}

\section{Annotation codes}
\label{sec:ap_codes}
Here, we report the codes used to label the ads.
\begin{squishlist}
\item EU: national sovereignty, Brexit, unity, power balance
\item Localism: separatism, autonomy
\item Foreign policy: peace, sanctions
Institutions: corruption, state reforms, church, police, banks, military
\item Security: law \& order
\item Environment: animal welfare, emissions, urban development, climate change, technology development, ecology
\item Human rights: LGBTQ+, women's rights, freedom of speech, digital rights, inclusivity, hate speech, voting
\item Immigration: relocation, integration, illegal immigration, racism
\item Economy: austerity, taxes, corporate tax, personal tax, welfare, healthcare, investments, education, jobs, workers' rights, housing, urban development, rural development, wages, better bureaucracy, euro, inequality, socialism, innovation, free trade
\end{squishlist}

\section{Model-based demographic analysis}
\label{sec:ap_model_1}

To understand whether party-specific differences in demographics are significant when taken together, we build a machine learning model that distinguishes whether the ads comes from a party with a specific \target tag given its reached audience.
For each ad, we use as features the distribution over the Cartesian product of gender and age buckets, which results in $14$ different features.\footnote{The features originally sum to one, but we standardize their values along the columns, so linear dependency is not an issue (verified via VIF analysis).}
We use support vector machine (SVM) as the classifier for the task, and we train a separate model for each tag and country.

\begin{figure}[h]
    \centering
    \includegraphics[width=0.8\columnwidth]{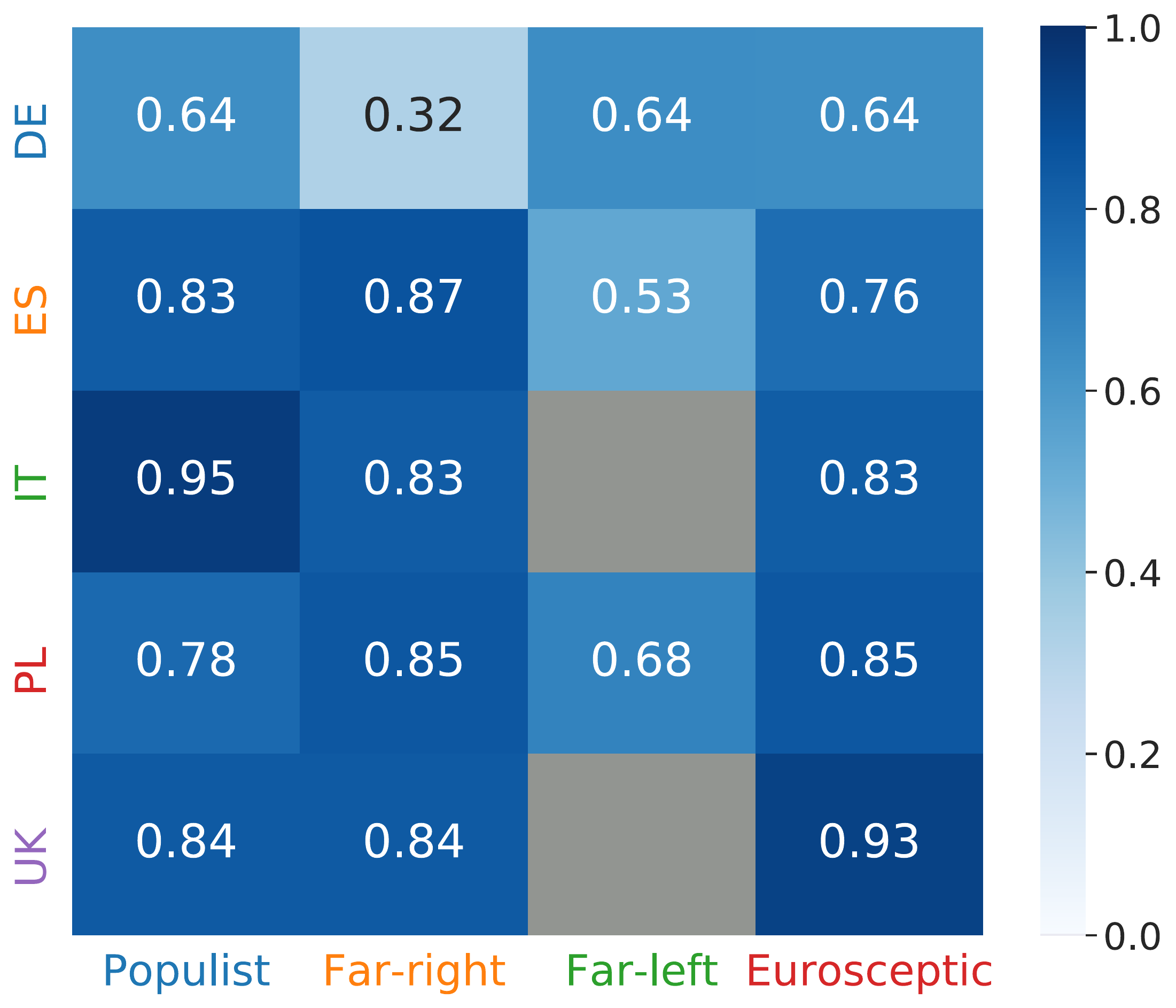}
    \caption{F1 score of the demographic classifier in 10-fold cross-validation (grey squares have fewer than 10 ads).}
    \label{fig:f1_demog_cval}
    \Description{F1 score of the demographic classifier in 10-fold cross-validation (grey squares have fewer than 10 ads)}
\end{figure}

\Cref{fig:f1_demog_cval} shows, for each country, the F1 score (harmonic mean between Precision and Recall), computed in stratified 10-fold cross-validation, of a classifier that distinguishes ads with a given tag from those without it based on the audience demographic.
In general, the demographic-based classifier manages to distinguish populist, far-right, and eurosceptic ads with a sufficient level of accuracy (F1 $\geq 0.64$).
Conversely, far-left ads are less distinguishable from their counterparts.
Finally, far-right ads in Germany are harder to distinguish, despite their very different average audience (see \Cref{fig:wass_odds} and \Cref{fig:age_distribution}).
The main reason is that they represent a small minority of the total amount of ads in the country, all from a single party (20 from AfD -- Alternative f\"{u}r Deutschland, see \Cref{tab:ad_stats}).
However, when looking at the individual ads, we find a high variability in their reached audience: some tend to reach a younger population, while others are shown to older individuals.
This variability makes learning a consistent distinguishing pattern almost impossible.
Overall, \target ads seem to be relatively easy to identify given the demographics of their audiences, within each country, as long as they do not constitute too small of a minority.

\section{Content Analysis}
\label{sec:ap_content}

\begin{figure}[H]
    \centering
    \includegraphics[width=\columnwidth]{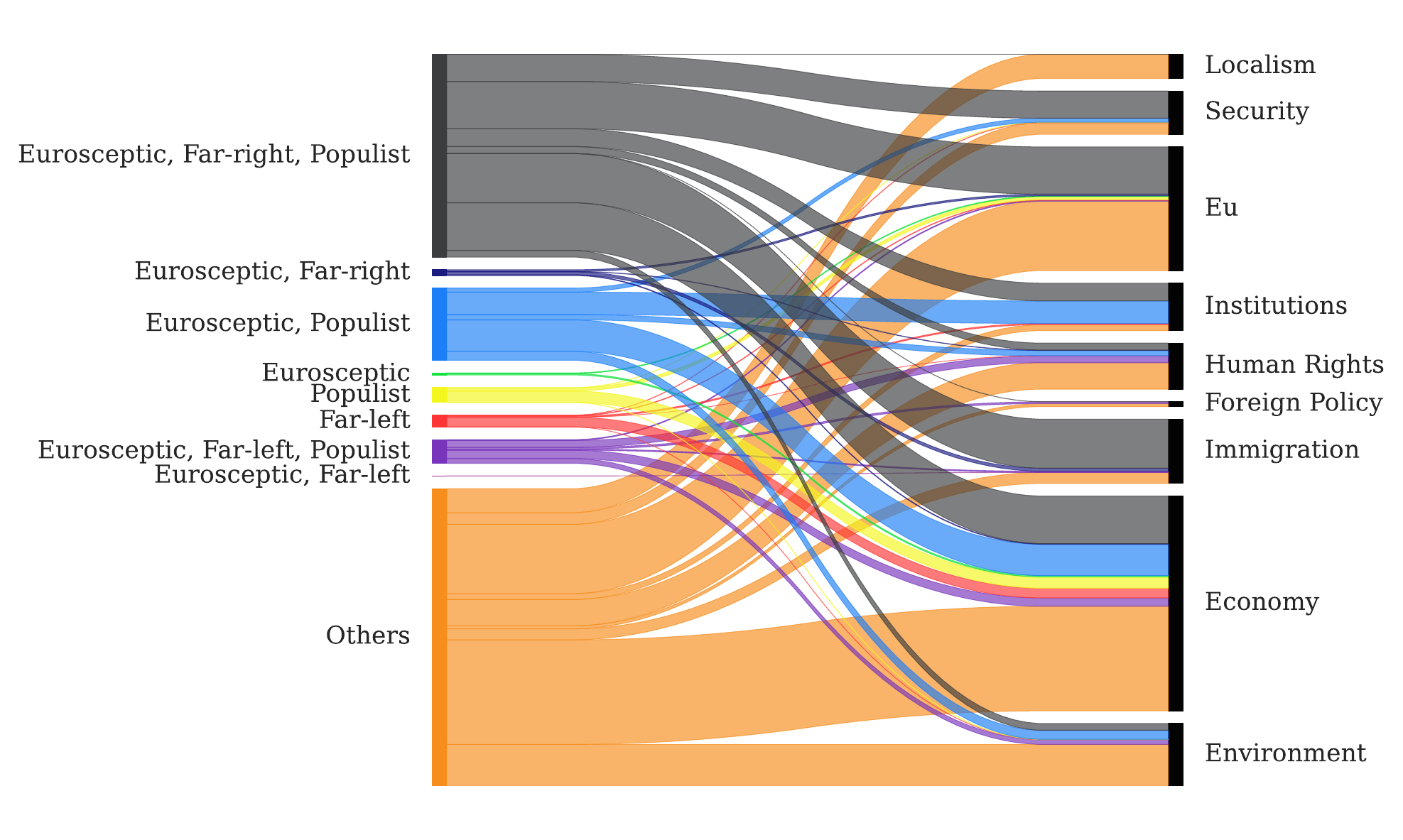}
    \caption{Sankey diagram showing how much each issue (on the right) is mentioned by Facebook ads of parties with a specific combination of \target tags (on the left). We indicate with ``others'' all parties that do not have any \target tag. Each ad is weighted by its number of impressions.}
    \label{fig:sankey}
    \Description{Sankey diagram showing how much each issue (on the right) is mentioned by Facebook ads of parties with a specific combination of PopuList tags (on the left). We indicate with ``others'' all parties that do not have any PopuList tag. Each ad is weighted by its number of impressions.}
\end{figure}

\section{Cross-country Demographic Similarity}
\label{sec:ap_demo_similarity}

\begin{figure}[b]
    \centering
    \includegraphics[width=\columnwidth]{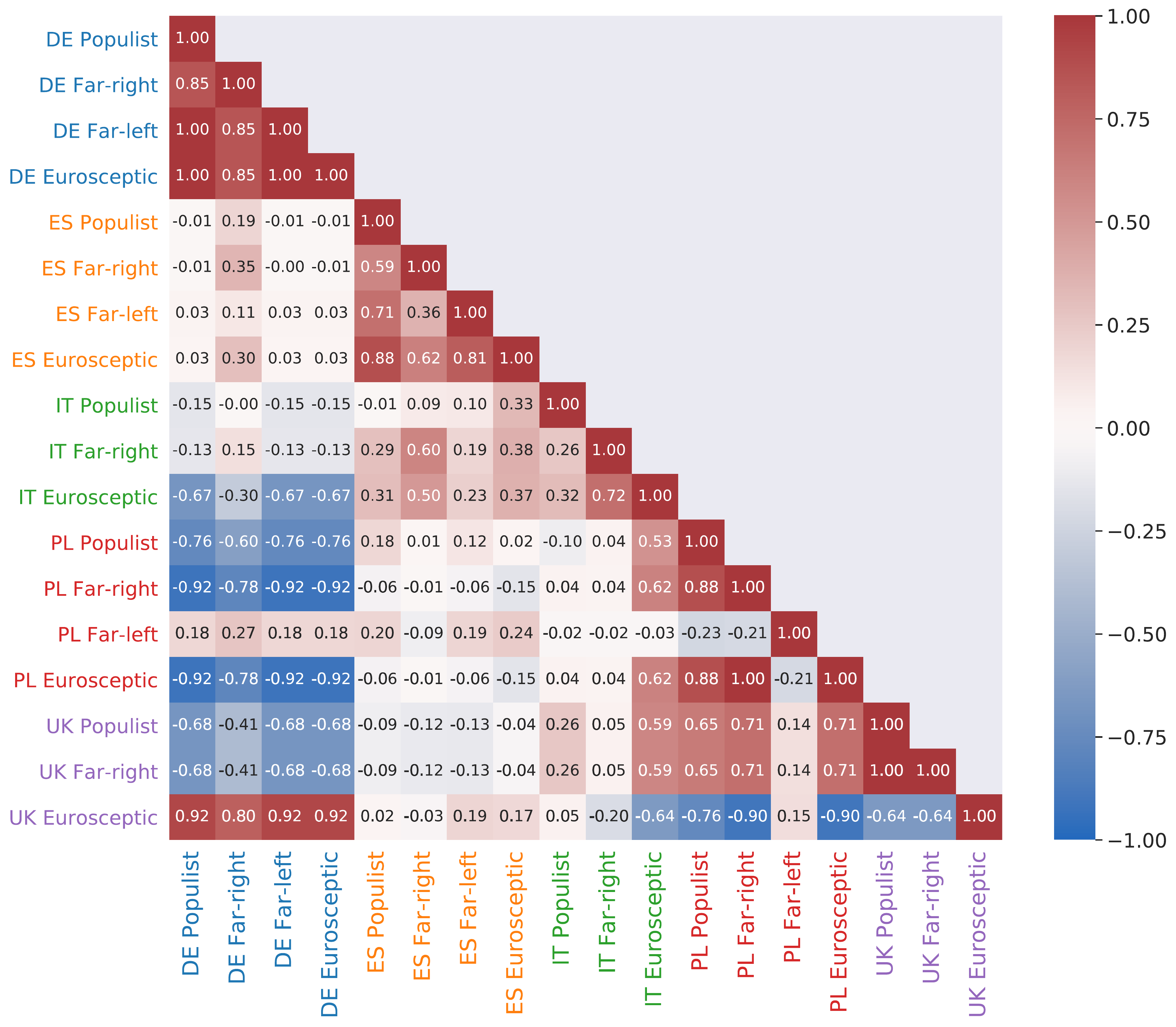}
    \caption{Cosine similarity of SVM coefficients for demographic classifier}
    \label{fig:similarity_demog}
    \Description{Cosine similarity of SVM coefficients for demographic classifier}
\end{figure}

\Cref{fig:similarity_demog} reports the cosine similarity of the SVM coefficients of each classifier across countries and tags.
The diagonal shows country-specific effects which results in blocks of high similarity within each country, with the exception of far-left, that often emerges as an outlier.
Some of this effect can clearly be attributed to the overlap between tags (mostly populist, far-right, and eurosceptic within the same country.

\begin{figure}[H]
    \centering
    \includegraphics[width=0.8\columnwidth]{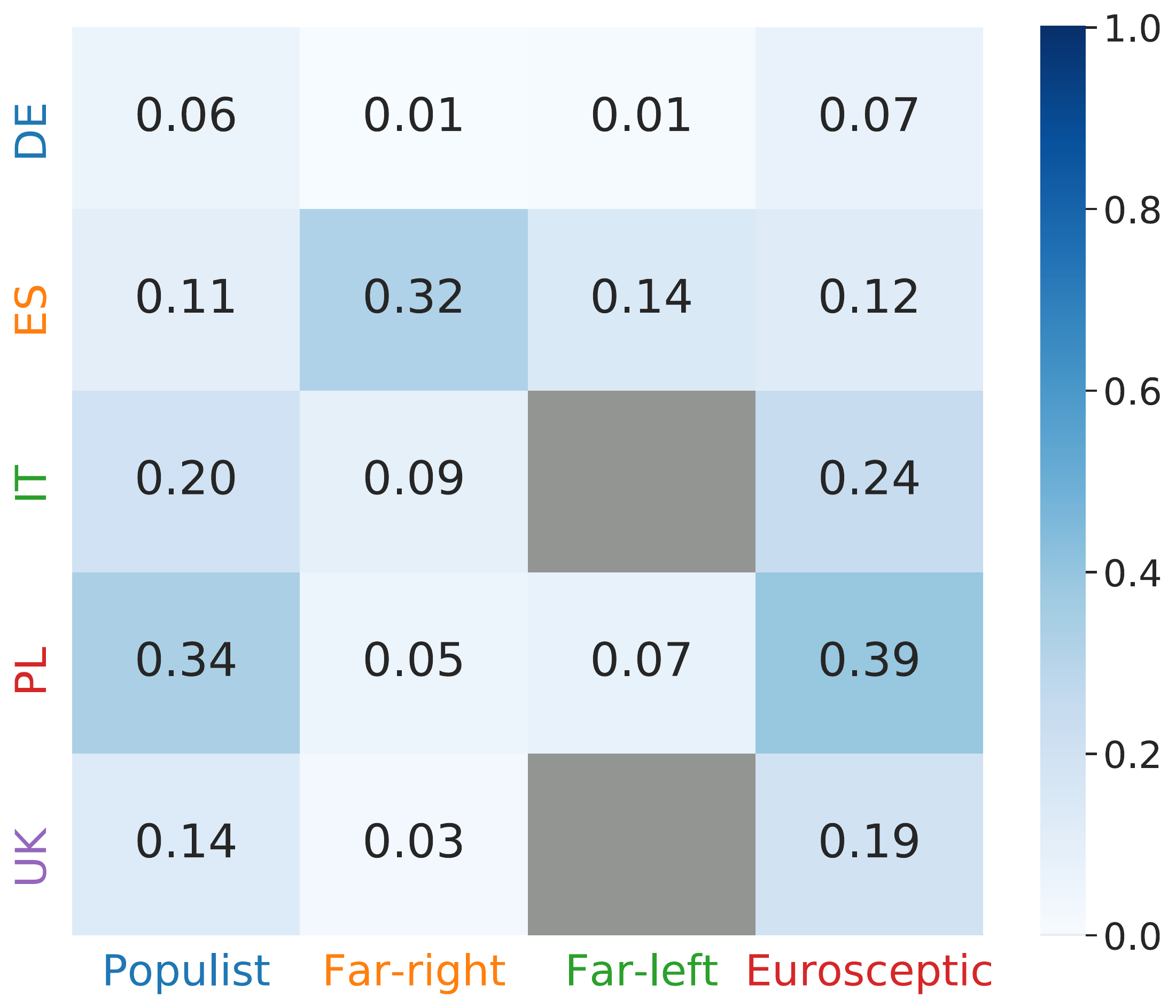}
    \caption{F1 score for domain adaptation task using demographic information. Larger values indicate greater similarities between parties with that tag in a given country, and parties with that tag in the other countries.}
    \label{fig:f1_demog_domain_adaptation}
    \Description{F1 score for domain adaptation task using demographic information. Larger values indicate greater similarities between parties with that tag in a given country, and parties with that tag in the other countries.}
\end{figure}

We examine the similarities across the audiences of ads from different countries by formulating a \emph{domain adaptation} task:
for each tag, we test a classifier for each country that is trained on all the ads from the other countries.
\Cref{fig:f1_demog_domain_adaptation} reports the F1 score for each domain adaptation classifier when trained on demographic data.
Most of the F1 scores are quite low.
This finding suggests that for the studied tags, differences across European countries are more prominent than any similarities in their demographics.
This result is consistent with \Cref{fig:age_distribution}, as the columns (i.e., tags) present substantial variability in their distributions.

%% file: main.bbl

\begin{thebibliography}{46}


\ifx \showCODEN    \undefined \def \showCODEN     #1{\unskip}     \fi
\ifx \showDOI      \undefined \def \showDOI       #1{#1}\fi
\ifx \showISBNx    \undefined \def \showISBNx     #1{\unskip}     \fi
\ifx \showISBNxiii \undefined \def \showISBNxiii  #1{\unskip}     \fi
\ifx \showISSN     \undefined \def \showISSN      #1{\unskip}     \fi
\ifx \showLCCN     \undefined \def \showLCCN      #1{\unskip}     \fi
\ifx \shownote     \undefined \def \shownote      #1{#1}          \fi
\ifx \showarticletitle \undefined \def \showarticletitle #1{#1}   \fi
\ifx \showURL      \undefined \def \showURL       {\relax}        \fi
\providecommand\bibfield[2]{#2}
\providecommand\bibinfo[2]{#2}
\providecommand\natexlab[1]{#1}
\providecommand\showeprint[2][]{arXiv:#2}

\bibitem[\protect\citeauthoryear{Abts and Rummens}{Abts and Rummens}{2007}]%
        {abts2007populism}
\bibfield{author}{\bibinfo{person}{Koen Abts} {and} \bibinfo{person}{Stefan
  Rummens}.} \bibinfo{year}{2007}\natexlab{}.
\newblock \showarticletitle{Populism versus democracy}.
\newblock \bibinfo{journal}{\emph{Political studies}} \bibinfo{volume}{55},
  \bibinfo{number}{2} (\bibinfo{year}{2007}), \bibinfo{pages}{405--424}.
\newblock


\bibitem[\protect\citeauthoryear{Albertazzi and McDonnell}{Albertazzi and
  McDonnell}{2008}]%
        {albertazzi2008introduction}
\bibfield{author}{\bibinfo{person}{Daniele Albertazzi} {and}
  \bibinfo{person}{Duncan McDonnell}.} \bibinfo{year}{2008}\natexlab{}.
\newblock \showarticletitle{Introduction: The sceptre and the spectre}.
\newblock In \bibinfo{booktitle}{\emph{Twenty-first century populism}}.
  \bibinfo{publisher}{Springer}, \bibinfo{pages}{1--11}.
\newblock


\bibitem[\protect\citeauthoryear{Ali, Sapiezynski, Bogen, Korolova, Mislove,
  and Rieke}{Ali et~al\mbox{.}}{2019}]%
        {ali2019discrimination}
\bibfield{author}{\bibinfo{person}{Muhammad Ali}, \bibinfo{person}{Piotr
  Sapiezynski}, \bibinfo{person}{Miranda Bogen}, \bibinfo{person}{Aleksandra
  Korolova}, \bibinfo{person}{Alan Mislove}, {and} \bibinfo{person}{Aaron
  Rieke}.} \bibinfo{year}{2019}\natexlab{}.
\newblock \showarticletitle{Discrimination through optimization: How Facebook's
  Ad delivery can lead to biased outcomes}. In \bibinfo{booktitle}{\emph{CSCW:
  Proceedings of the ACM on Human-Computer Interaction}},
  Vol.~\bibinfo{volume}{3}. \bibinfo{pages}{1--30}.
\newblock


\bibitem[\protect\citeauthoryear{Ali, Sapiezynski, Korolova, Mislove, and
  Rieke}{Ali et~al\mbox{.}}{2021}]%
        {ali2021ad}
\bibfield{author}{\bibinfo{person}{Muhammad Ali}, \bibinfo{person}{Piotr
  Sapiezynski}, \bibinfo{person}{Aleksandra Korolova}, \bibinfo{person}{Alan
  Mislove}, {and} \bibinfo{person}{Aaron Rieke}.}
  \bibinfo{year}{2021}\natexlab{}.
\newblock \showarticletitle{Ad Delivery Algorithms: The Hidden Arbiters of
  Political Messaging}. In \bibinfo{booktitle}{\emph{Proceedings of the 14th
  ACM International Conference on Web Search and Data Mining}}.
  \bibinfo{pages}{13--21}.
\newblock


\bibitem[\protect\citeauthoryear{Araujo, Mejova, Weber, and Benevenuto}{Araujo
  et~al\mbox{.}}{2017}]%
        {araujo2017using}
\bibfield{author}{\bibinfo{person}{Matheus Araujo}, \bibinfo{person}{Yelena
  Mejova}, \bibinfo{person}{Ingmar Weber}, {and} \bibinfo{person}{Fabricio
  Benevenuto}.} \bibinfo{year}{2017}\natexlab{}.
\newblock \showarticletitle{Using Facebook ads audiences for global lifestyle
  disease surveillance: Promises and limitations}. In
  \bibinfo{booktitle}{\emph{Proceedings of the 2017 ACM on Web science
  conference}}. \bibinfo{pages}{253--257}.
\newblock


\bibitem[\protect\citeauthoryear{Bennett and Seyis}{Bennett and Seyis}{2021}]%
        {bennett2021online}
\bibfield{author}{\bibinfo{person}{Andrew Bennett} {and} \bibinfo{person}{Didem
  Seyis}.} \bibinfo{year}{2021}\natexlab{}.
\newblock \showarticletitle{The Online Market's Invisible Hand: Internet Media
  and Rising Populism}.
\newblock \bibinfo{journal}{\emph{Political Studies}} (\bibinfo{year}{2021}),
  \bibinfo{pages}{00323217211033230}.
\newblock


\bibitem[\protect\citeauthoryear{Cano~Or{\'o}n, Calvo, L{\'o}pez~Garc{\'\i}a,
  and Baviera}{Cano~Or{\'o}n et~al\mbox{.}}{2021}]%
        {cano2021disinformation}
\bibfield{author}{\bibinfo{person}{Lorena Cano~Or{\'o}n},
  \bibinfo{person}{Dafne Calvo}, \bibinfo{person}{Guillermo
  L{\'o}pez~Garc{\'\i}a}, {and} \bibinfo{person}{Tom{\'a}s Baviera}.}
  \bibinfo{year}{2021}\natexlab{}.
\newblock \showarticletitle{Disinformation in Facebook Ads in the 2019 Spanish
  general election campaigns}.
\newblock \bibinfo{journal}{\emph{Media And Communication, 2021, vol. 9, num.
  1, p. 217-228}} (\bibinfo{year}{2021}).
\newblock


\bibitem[\protect\citeauthoryear{Capozzi, De~Francisci~Morales, Mejova, Monti,
  Panisson, and Paolotti}{Capozzi et~al\mbox{.}}{2021}]%
        {capozzi2021clandestino}
\bibfield{author}{\bibinfo{person}{Arthur Capozzi}, \bibinfo{person}{Gianmarco
  De~Francisci~Morales}, \bibinfo{person}{Yelena Mejova},
  \bibinfo{person}{Corrado Monti}, \bibinfo{person}{Andr{\'e} Panisson}, {and}
  \bibinfo{person}{Daniela Paolotti}.} \bibinfo{year}{2021}\natexlab{}.
\newblock \showarticletitle{Clandestino or Rifugiato? Anti-immigration Facebook
  Ad Targeting in Italy}. In \bibinfo{booktitle}{\emph{Proceedings of the 2021
  CHI Conference on Human Factors in Computing Systems}}.
  \bibinfo{pages}{1--15}.
\newblock


\bibitem[\protect\citeauthoryear{Capozzi, Francisci~Morales, Mejova, Monti,
  Panisson, and Paolotti}{Capozzi et~al\mbox{.}}{2020}]%
        {capozzi2020facebook}
\bibfield{author}{\bibinfo{person}{Arthur Capozzi},
  \bibinfo{person}{Gianmarco~De Francisci~Morales}, \bibinfo{person}{Yelena
  Mejova}, \bibinfo{person}{Corrado Monti}, \bibinfo{person}{Andr{\'e}
  Panisson}, {and} \bibinfo{person}{Daniela Paolotti}.}
  \bibinfo{year}{2020}\natexlab{}.
\newblock \showarticletitle{Facebook Ads: Politics of Migration in Italy}. In
  \bibinfo{booktitle}{\emph{International Conference on Social Informatics}}.
  Springer, \bibinfo{pages}{43--57}.
\newblock


\bibitem[\protect\citeauthoryear{Cinelli, De~Francisci~Morales, Galeazzi,
  Quattrociocchi, and Starnini}{Cinelli et~al\mbox{.}}{2021}]%
        {cinelli2021echo}
\bibfield{author}{\bibinfo{person}{Matteo Cinelli}, \bibinfo{person}{Gianmarco
  De~Francisci~Morales}, \bibinfo{person}{Alessandro Galeazzi},
  \bibinfo{person}{Walter Quattrociocchi}, {and} \bibinfo{person}{Michele
  Starnini}.} \bibinfo{year}{2021}\natexlab{}.
\newblock \showarticletitle{The {Echo} {Chamber} {Effect} on {Social} {Media}}.
\newblock \bibinfo{journal}{\emph{PNAS: Proceedings of the National Academy of
  Sciences}} \bibinfo{volume}{118}, \bibinfo{number}{9} (\bibinfo{year}{2021}),
  \bibinfo{pages}{e2023301118}.
\newblock
\urldef\tempurl%
\url{https://doi.org/10.1073/pnas.2023301118}
\showDOI{\tempurl}


\bibitem[\protect\citeauthoryear{Colantone and Stanig}{Colantone and
  Stanig}{2018}]%
        {colantone2018trade}
\bibfield{author}{\bibinfo{person}{Italo Colantone} {and}
  \bibinfo{person}{Piero Stanig}.} \bibinfo{year}{2018}\natexlab{}.
\newblock \showarticletitle{The trade origins of economic nationalism: Import
  competition and voting behavior in Western Europe}.
\newblock \bibinfo{journal}{\emph{American Journal of Political Science}}
  \bibinfo{volume}{62}, \bibinfo{number}{4} (\bibinfo{year}{2018}),
  \bibinfo{pages}{936--953}.
\newblock


\bibitem[\protect\citeauthoryear{Dinas, Matakos, Xefteris, and
  Hangartner}{Dinas et~al\mbox{.}}{2019}]%
        {dinas2019waking}
\bibfield{author}{\bibinfo{person}{Elias Dinas}, \bibinfo{person}{Konstantinos
  Matakos}, \bibinfo{person}{Dimitrios Xefteris}, {and}
  \bibinfo{person}{Dominik Hangartner}.} \bibinfo{year}{2019}\natexlab{}.
\newblock \showarticletitle{Waking up the golden dawn: does exposure to the
  refugee crisis increase support for extreme-right parties?}
\newblock \bibinfo{journal}{\emph{Political analysis}} \bibinfo{volume}{27},
  \bibinfo{number}{2} (\bibinfo{year}{2019}), \bibinfo{pages}{244--254}.
\newblock


\bibitem[\protect\citeauthoryear{Dobber, {\'O}~Fathaigh, and
  Zuiderveen~Borgesius}{Dobber et~al\mbox{.}}{2019}]%
        {dobber2019regulation}
\bibfield{author}{\bibinfo{person}{Tom Dobber}, \bibinfo{person}{Ronan
  {\'O}~Fathaigh}, {and} \bibinfo{person}{Frederik Zuiderveen~Borgesius}.}
  \bibinfo{year}{2019}\natexlab{}.
\newblock \showarticletitle{The regulation of online political micro-targeting
  in Europe}.
\newblock \bibinfo{journal}{\emph{Internet Policy Review}} \bibinfo{volume}{8},
  \bibinfo{number}{4} (\bibinfo{year}{2019}).
\newblock


\bibitem[\protect\citeauthoryear{Dommett and Power}{Dommett and Power}{2019}]%
        {dommett2019political}
\bibfield{author}{\bibinfo{person}{Katharine Dommett} {and}
  \bibinfo{person}{Sam Power}.} \bibinfo{year}{2019}\natexlab{}.
\newblock \showarticletitle{The political economy of Facebook advertising:
  Election spending, regulation and targeting online}.
\newblock \bibinfo{journal}{\emph{The Political Quarterly}}
  \bibinfo{volume}{90}, \bibinfo{number}{2} (\bibinfo{year}{2019}),
  \bibinfo{pages}{257--265}.
\newblock


\bibitem[\protect\citeauthoryear{Edelson, Lauinger, and McCoy}{Edelson
  et~al\mbox{.}}{2020}]%
        {edelson2020security}
\bibfield{author}{\bibinfo{person}{Laura Edelson}, \bibinfo{person}{Tobias
  Lauinger}, {and} \bibinfo{person}{Damon McCoy}.}
  \bibinfo{year}{2020}\natexlab{}.
\newblock \showarticletitle{A security analysis of the Facebook ad library}. In
  \bibinfo{booktitle}{\emph{2020 IEEE Symposium on Security and Privacy (SP)}}.
  IEEE, \bibinfo{pages}{661--678}.
\newblock


\bibitem[\protect\citeauthoryear{Entous, Timberg, and Dwoskin}{Entous
  et~al\mbox{.}}{2017}]%
        {entous2017russian}
\bibfield{author}{\bibinfo{person}{Adam Entous}, \bibinfo{person}{Craig
  Timberg}, {and} \bibinfo{person}{Elizabeth Dwoskin}.}
  \bibinfo{year}{2017}\natexlab{}.
\newblock \bibinfo{title}{Russian operatives used Facebook ads to exploit
  America's racial and religious divisions}.
\newblock
\newblock


\bibitem[\protect\citeauthoryear{Ernst, Esser, Blassnig, and Engesser}{Ernst
  et~al\mbox{.}}{2019}]%
        {ernst2019favorable}
\bibfield{author}{\bibinfo{person}{Nicole Ernst}, \bibinfo{person}{Frank
  Esser}, \bibinfo{person}{Sina Blassnig}, {and} \bibinfo{person}{Sven
  Engesser}.} \bibinfo{year}{2019}\natexlab{}.
\newblock \showarticletitle{Favorable opportunity structures for populist
  communication: Comparing different types of politicians and issues in social
  media, television and the press}.
\newblock \bibinfo{journal}{\emph{The International Journal of Press/Politics}}
  \bibinfo{volume}{24}, \bibinfo{number}{2} (\bibinfo{year}{2019}),
  \bibinfo{pages}{165--188}.
\newblock


\bibitem[\protect\citeauthoryear{Esser, St{\k{e}}pi{\'n}ska, and Hopmann}{Esser
  et~al\mbox{.}}{2017}]%
        {esser201728}
\bibfield{author}{\bibinfo{person}{Frank Esser}, \bibinfo{person}{Agnieszka
  St{\k{e}}pi{\'n}ska}, {and} \bibinfo{person}{David~Nicolas Hopmann}.}
  \bibinfo{year}{2017}\natexlab{}.
\newblock \showarticletitle{28. Populism and the Media. Cross-National Findings
  and Perspectives}.
\newblock \bibinfo{journal}{\emph{T. Aalberg, F. Esser, C. Reinemann, J.
  Str{\"o}mb{\"a}ck \& C. d. Vreese (Eds.), Populist political communication in
  Europe}} (\bibinfo{year}{2017}), \bibinfo{pages}{365--380}.
\newblock


\bibitem[\protect\citeauthoryear{Gerbaudo}{Gerbaudo}{2018}]%
        {gerbaudo2018social}
\bibfield{author}{\bibinfo{person}{Paolo Gerbaudo}.}
  \bibinfo{year}{2018}\natexlab{}.
\newblock \showarticletitle{Social media and populism: an elective affinity?}
\newblock \bibinfo{journal}{\emph{Media, culture \& society}}
  \bibinfo{volume}{40}, \bibinfo{number}{5} (\bibinfo{year}{2018}),
  \bibinfo{pages}{745--753}.
\newblock


\bibitem[\protect\citeauthoryear{Guiso, Herrera, Morelli, and Sonno}{Guiso
  et~al\mbox{.}}{2019}]%
        {guiso2019global}
\bibfield{author}{\bibinfo{person}{Luigi Guiso}, \bibinfo{person}{Helios
  Herrera}, \bibinfo{person}{Massimo Morelli}, {and} \bibinfo{person}{Tommaso
  Sonno}.} \bibinfo{year}{2019}\natexlab{}.
\newblock \showarticletitle{Global crises and populism: the role of Eurozone
  institutions}.
\newblock \bibinfo{journal}{\emph{Economic Policy}} \bibinfo{volume}{34},
  \bibinfo{number}{97} (\bibinfo{year}{2019}), \bibinfo{pages}{95--139}.
\newblock


\bibitem[\protect\citeauthoryear{Holroyd}{Holroyd}{2021}]%
        {holroyd}
\bibfield{author}{\bibinfo{person}{Matthew Holroyd}.}
  \bibinfo{year}{2021}\natexlab{}.
\newblock \bibinfo{title}{German election: Who has spent the most on Facebook
  advertising?}
\newblock
  \bibinfo{howpublished}{https://www.euronews.com/2021/09/25/german-election-who-has-spent-the-most-on-facebook-advertising}.
\newblock


\bibitem[\protect\citeauthoryear{Hooghe and Marks}{Hooghe and Marks}{2018}]%
        {hooghe2018cleavage}
\bibfield{author}{\bibinfo{person}{Liesbet Hooghe} {and} \bibinfo{person}{Gary
  Marks}.} \bibinfo{year}{2018}\natexlab{}.
\newblock \showarticletitle{Cleavage theory meets Europe's crises: Lipset,
  Rokkan, and the transnational cleavage}.
\newblock \bibinfo{journal}{\emph{Journal of European public policy}}
  \bibinfo{volume}{25}, \bibinfo{number}{1} (\bibinfo{year}{2018}),
  \bibinfo{pages}{109--135}.
\newblock


\bibitem[\protect\citeauthoryear{Karpf}{Karpf}{2016}]%
        {karpf2016analytic}
\bibfield{author}{\bibinfo{person}{David Karpf}.}
  \bibinfo{year}{2016}\natexlab{}.
\newblock \bibinfo{booktitle}{\emph{Analytic activism: Digital listening and
  the new political strategy}}.
\newblock \bibinfo{publisher}{Oxford University Press}.
\newblock


\bibitem[\protect\citeauthoryear{Koc-Michalska, Lilleker, Michalski, Gibson,
  and Zajac}{Koc-Michalska et~al\mbox{.}}{2021}]%
        {koc2021facebook}
\bibfield{author}{\bibinfo{person}{Karolina Koc-Michalska},
  \bibinfo{person}{Darren~G Lilleker}, \bibinfo{person}{Tomasz Michalski},
  \bibinfo{person}{Rachel Gibson}, {and} \bibinfo{person}{Jan~M Zajac}.}
  \bibinfo{year}{2021}\natexlab{}.
\newblock \showarticletitle{Facebook affordances and citizen engagement during
  elections: European political parties and their benefit from online
  strategies?}
\newblock \bibinfo{journal}{\emph{Journal of Information Technology \&
  Politics}} \bibinfo{volume}{18}, \bibinfo{number}{2} (\bibinfo{year}{2021}),
  \bibinfo{pages}{180--193}.
\newblock


\bibitem[\protect\citeauthoryear{Kriesi and Pappas}{Kriesi and Pappas}{2015}]%
        {kriesi2015european}
\bibfield{author}{\bibinfo{person}{Hanspeter Kriesi} {and}
  \bibinfo{person}{Takis~S Pappas}.} \bibinfo{year}{2015}\natexlab{}.
\newblock \bibinfo{booktitle}{\emph{European populism in the shadow of the
  great recession}}.
\newblock \bibinfo{publisher}{Ecpr Press Colchester}.
\newblock


\bibitem[\protect\citeauthoryear{Kruikemeier, Sezgin, and Boerman}{Kruikemeier
  et~al\mbox{.}}{2016}]%
        {kruikemeier2016political}
\bibfield{author}{\bibinfo{person}{Sanne Kruikemeier}, \bibinfo{person}{Minem
  Sezgin}, {and} \bibinfo{person}{Sophie~C Boerman}.}
  \bibinfo{year}{2016}\natexlab{}.
\newblock \showarticletitle{Political microtargeting: relationship between
  personalized advertising on Facebook and voters' responses}.
\newblock \bibinfo{journal}{\emph{Cyberpsychology, Behavior, and Social
  Networking}} \bibinfo{volume}{19}, \bibinfo{number}{6}
  (\bibinfo{year}{2016}), \bibinfo{pages}{367--372}.
\newblock


\bibitem[\protect\citeauthoryear{Laclau}{Laclau}{2005}]%
        {laclau2005populism}
\bibfield{author}{\bibinfo{person}{Ernesto Laclau}.}
  \bibinfo{year}{2005}\natexlab{}.
\newblock \showarticletitle{Populism: What’s in a Name?}
\newblock \bibinfo{journal}{\emph{Populism and the Mirror of Democracy}}
  \bibinfo{volume}{48} (\bibinfo{year}{2005}).
\newblock


\bibitem[\protect\citeauthoryear{Linneberg and Korsgaard}{Linneberg and
  Korsgaard}{2019}]%
        {linneberg2019coding}
\bibfield{author}{\bibinfo{person}{Mai~Skjott Linneberg} {and}
  \bibinfo{person}{Steffen Korsgaard}.} \bibinfo{year}{2019}\natexlab{}.
\newblock \showarticletitle{Coding qualitative data: A synthesis guiding the
  novice}.
\newblock \bibinfo{journal}{\emph{Qualitative research journal}}
  (\bibinfo{year}{2019}).
\newblock


\bibitem[\protect\citeauthoryear{Matias, Hounsel, and Feamster}{Matias
  et~al\mbox{.}}{2022}]%
        {matias2022software}
\bibfield{author}{\bibinfo{person}{J~Nathan Matias}, \bibinfo{person}{Austin
  Hounsel}, {and} \bibinfo{person}{Nick Feamster}.}
  \bibinfo{year}{2022}\natexlab{}.
\newblock \showarticletitle{Software-Supported Audits of Decision-Making
  Systems: Testing Google and Facebook's Political Advertising Policies}. In
  \bibinfo{booktitle}{\emph{CSCW: Proceedings of the ACM on Human-Computer
  Interaction}}, Vol.~\bibinfo{volume}{6}. \bibinfo{pages}{1--19}.
\newblock


\bibitem[\protect\citeauthoryear{Mejova and Kalimeri}{Mejova and
  Kalimeri}{2020}]%
        {mejova2020covid}
\bibfield{author}{\bibinfo{person}{Yelena Mejova} {and}
  \bibinfo{person}{Kyriaki Kalimeri}.} \bibinfo{year}{2020}\natexlab{}.
\newblock \showarticletitle{COVID-19 on Facebook ads: competing agendas around
  a public health crisis}. In \bibinfo{booktitle}{\emph{Proceedings of the 3rd
  ACM SIGCAS Conference on Computing and Sustainable Societies}}.
  \bibinfo{pages}{22--31}.
\newblock


\bibitem[\protect\citeauthoryear{Moffitt}{Moffitt}{2018}]%
        {moffitt2018populism}
\bibfield{author}{\bibinfo{person}{Benjamin Moffitt}.}
  \bibinfo{year}{2018}\natexlab{}.
\newblock \showarticletitle{The populism/anti-populism divide in Western
  Europe}.
\newblock \bibinfo{journal}{\emph{Democratic Theory}} \bibinfo{volume}{5},
  \bibinfo{number}{2} (\bibinfo{year}{2018}), \bibinfo{pages}{1--16}.
\newblock


\bibitem[\protect\citeauthoryear{Mudde and Kaltwasser}{Mudde and
  Kaltwasser}{2017}]%
        {mudde2017populism}
\bibfield{author}{\bibinfo{person}{Cas Mudde} {and}
  \bibinfo{person}{Crist{\'o}bal~Rovira Kaltwasser}.}
  \bibinfo{year}{2017}\natexlab{}.
\newblock \bibinfo{booktitle}{\emph{Populism: A very short introduction}}.
\newblock \bibinfo{publisher}{Oxford University Press}.
\newblock


\bibitem[\protect\citeauthoryear{Mudde, Kaltwasser, et~al\mbox{.}}{Mudde
  et~al\mbox{.}}{2012}]%
        {mudde2012populism}
\bibfield{author}{\bibinfo{person}{Cas Mudde},
  \bibinfo{person}{Crist{\'o}bal~Rovira Kaltwasser}, {et~al\mbox{.}}}
  \bibinfo{year}{2012}\natexlab{}.
\newblock \showarticletitle{Populism and (liberal) democracy: a framework for
  analysis}.
\newblock \bibinfo{journal}{\emph{Populism in Europe and the Americas: Threat
  or corrective for democracy}} \bibinfo{volume}{1}, \bibinfo{number}{5}
  (\bibinfo{year}{2012}).
\newblock


\bibitem[\protect\citeauthoryear{Noury and Roland}{Noury and Roland}{2020}]%
        {noury2020identity}
\bibfield{author}{\bibinfo{person}{Abdul Noury} {and} \bibinfo{person}{Gerard
  Roland}.} \bibinfo{year}{2020}\natexlab{}.
\newblock \showarticletitle{Identity politics and populism in Europe}.
\newblock \bibinfo{journal}{\emph{Annual Review of Political Science}}
  \bibinfo{volume}{23} (\bibinfo{year}{2020}), \bibinfo{pages}{421--439}.
\newblock


\bibitem[\protect\citeauthoryear{Peck}{Peck}{2014}]%
        {peck2014you}
\bibfield{author}{\bibinfo{person}{Reece Peck}.}
  \bibinfo{year}{2014}\natexlab{}.
\newblock \showarticletitle{`You say rich, I say job creator': how Fox News
  framed the Great Recession through the moral discourse of producerism}.
\newblock \bibinfo{journal}{\emph{Media, Culture \& Society}}
  \bibinfo{volume}{36}, \bibinfo{number}{4} (\bibinfo{year}{2014}),
  \bibinfo{pages}{526--535}.
\newblock


\bibitem[\protect\citeauthoryear{Rooduijn, Van~Kessel, Froio, Pirro, De~Lange,
  Halikiopoulou, Lewis, Mudde, and Taggart}{Rooduijn et~al\mbox{.}}{2019}]%
        {rooduijn2019populist}
\bibfield{author}{\bibinfo{person}{Matthijs Rooduijn}, \bibinfo{person}{Stijn
  Van~Kessel}, \bibinfo{person}{Caterina Froio}, \bibinfo{person}{Andrea
  Pirro}, \bibinfo{person}{Sarah De~Lange}, \bibinfo{person}{Daphne
  Halikiopoulou}, \bibinfo{person}{Paul Lewis}, \bibinfo{person}{Cas Mudde},
  {and} \bibinfo{person}{Paul Taggart}.} \bibinfo{year}{2019}\natexlab{}.
\newblock \showarticletitle{The PopuList: An overview of populist, far right,
  far left and Eurosceptic parties in Europe}.
\newblock  (\bibinfo{year}{2019}).
\newblock


\bibitem[\protect\citeauthoryear{Rydgren}{Rydgren}{2018}]%
        {rydgren2018oxford}
\bibfield{author}{\bibinfo{person}{Jens Rydgren}.}
  \bibinfo{year}{2018}\natexlab{}.
\newblock \bibinfo{booktitle}{\emph{The Oxford handbook of the radical right}}.
\newblock \bibinfo{publisher}{Oxford University Press}.
\newblock


\bibitem[\protect\citeauthoryear{Sosnovik and Goga}{Sosnovik and Goga}{2021}]%
        {sosnovik2021understanding}
\bibfield{author}{\bibinfo{person}{Vera Sosnovik} {and} \bibinfo{person}{Oana
  Goga}.} \bibinfo{year}{2021}\natexlab{}.
\newblock \showarticletitle{Understanding the complexity of detecting political
  ads}. In \bibinfo{booktitle}{\emph{Proceedings of the Web Conference 2021}}.
  \bibinfo{pages}{2002--2013}.
\newblock


\bibitem[\protect\citeauthoryear{Speicher, Ali, Venkatadri, Ribeiro,
  Arvanitakis, Benevenuto, Gummadi, Loiseau, and Mislove}{Speicher
  et~al\mbox{.}}{2018}]%
        {speicher2018potential}
\bibfield{author}{\bibinfo{person}{Till Speicher}, \bibinfo{person}{Muhammad
  Ali}, \bibinfo{person}{Giridhari Venkatadri}, \bibinfo{person}{Filipe
  Ribeiro}, \bibinfo{person}{George Arvanitakis},
  \bibinfo{person}{Fabr{\'\i}cio Benevenuto}, \bibinfo{person}{Krishna
  Gummadi}, \bibinfo{person}{Patrick Loiseau}, {and} \bibinfo{person}{Alan
  Mislove}.} \bibinfo{year}{2018}\natexlab{}.
\newblock \showarticletitle{Potential for discrimination in online targeted
  advertising}. In \bibinfo{booktitle}{\emph{FAT 2018-Conference on Fairness,
  Accountability, and Transparency}}, Vol.~\bibinfo{volume}{81}.
  \bibinfo{pages}{1--15}.
\newblock


\bibitem[\protect\citeauthoryear{Stanley}{Stanley}{2008}]%
        {stanley2008thin}
\bibfield{author}{\bibinfo{person}{Ben Stanley}.}
  \bibinfo{year}{2008}\natexlab{}.
\newblock \showarticletitle{The thin ideology of populism}.
\newblock \bibinfo{journal}{\emph{Journal of political ideologies}}
  \bibinfo{volume}{13}, \bibinfo{number}{1} (\bibinfo{year}{2008}),
  \bibinfo{pages}{95--110}.
\newblock


\bibitem[\protect\citeauthoryear{Stockemer and Amengay}{Stockemer and
  Amengay}{2020}]%
        {stockemer20202019}
\bibfield{author}{\bibinfo{person}{Daniel Stockemer} {and}
  \bibinfo{person}{Abdelkarim Amengay}.} \bibinfo{year}{2020}\natexlab{}.
\newblock \showarticletitle{The 2019 Elections to the European Parliament: The
  Continuation of a Populist Wave but Not a Populist Tsunami}.
\newblock \bibinfo{journal}{\emph{JCMS: Journal of Common Market Studies}}
  \bibinfo{volume}{58} (\bibinfo{year}{2020}), \bibinfo{pages}{28--42}.
\newblock


\bibitem[\protect\citeauthoryear{Vachudova}{Vachudova}{2021}]%
        {vachudova2021populism}
\bibfield{author}{\bibinfo{person}{Milada~Anna Vachudova}.}
  \bibinfo{year}{2021}\natexlab{}.
\newblock \showarticletitle{Populism, Democracy, and Party System Change in
  Europe}.
\newblock \bibinfo{journal}{\emph{Annual Review of Political Science}}
  \bibinfo{volume}{24} (\bibinfo{year}{2021}), \bibinfo{pages}{471--498}.
\newblock


\bibitem[\protect\citeauthoryear{Vaidhyanathan}{Vaidhyanathan}{2017}]%
        {vaidhyanathan2017facebook}
\bibfield{author}{\bibinfo{person}{Siva Vaidhyanathan}.}
  \bibinfo{year}{2017}\natexlab{}.
\newblock \showarticletitle{Facebook wins, democracy loses}.
\newblock \bibinfo{journal}{\emph{New York Times}}  \bibinfo{volume}{8}
  (\bibinfo{year}{2017}).
\newblock


\bibitem[\protect\citeauthoryear{Williams}{Williams}{2016}]%
        {williams2016nigel}
\bibfield{author}{\bibinfo{person}{Zoe Williams}.}
  \bibinfo{year}{2016}\natexlab{}.
\newblock \bibinfo{title}{Nigel Farage’s victory speech was a triumph of poor
  taste and ugliness}.
\newblock
  \bibinfo{howpublished}{https://www.theguardian.com/commentisfree/2016/jun/24/nigel-farage-ugliness-bullet-fired}.
\newblock


\bibitem[\protect\citeauthoryear{Zagheni, Weber, and Gummadi}{Zagheni
  et~al\mbox{.}}{2017}]%
        {zagheni2017leveraging}
\bibfield{author}{\bibinfo{person}{Emilio Zagheni}, \bibinfo{person}{Ingmar
  Weber}, {and} \bibinfo{person}{Krishna Gummadi}.}
  \bibinfo{year}{2017}\natexlab{}.
\newblock \showarticletitle{Leveraging Facebook's advertising platform to
  monitor stocks of migrants}.
\newblock \bibinfo{journal}{\emph{Population and Development Review}}
  (\bibinfo{year}{2017}), \bibinfo{pages}{721--734}.
\newblock


\bibitem[\protect\citeauthoryear{Zulianello}{Zulianello}{2020}]%
        {zulianello2020varieties}
\bibfield{author}{\bibinfo{person}{Mattia Zulianello}.}
  \bibinfo{year}{2020}\natexlab{}.
\newblock \showarticletitle{Varieties of populist parties and party systems in
  Europe: From state-of-the-art to the application of a novel classification
  scheme to 66 parties in 33 countries}.
\newblock \bibinfo{journal}{\emph{Government and Opposition}}
  \bibinfo{volume}{55}, \bibinfo{number}{2} (\bibinfo{year}{2020}),
  \bibinfo{pages}{327--347}.
\newblock


\end{thebibliography}
